\documentclass[twocolumn, twocolappendix]{aastex631}

\usepackage{caption}
\usepackage{amsmath}
\usepackage{subcaption}
\usepackage{array} 

\usepackage{lipsum}


\begin{document}
\title{B-fields And dust in interstelLar fiLAments using Dust POLarization (BALLAD-POL): III. Grain alignment and disruption mechanisms in G34.43+0.24 using polarization observations from JCMT/POL-2}

\author[0009-0002-6171-9740]{Saikhom Pravash}
\affiliation{Indian Institute of Astrophysics, II Block, Koramangala, 560034, India; \href{mailto:saikhom.singh@iiap.res.in}{saikhom.singh@iiap.res.in}, \href{mailto:spravash11@gmail.com}{spravash11@gmail.com}}
\affiliation{Pondicherry University, R.V. Nagar, Kalapet, Puducherry, 605014, India}

\author[0000-0002-6386-2906]{Archana Soam}
\affiliation{Indian Institute of Astrophysics, II Block, Koramangala, 560034, India; \href{mailto:saikhom.singh@iiap.res.in}{saikhom.singh@iiap.res.in}, \href{mailto:spravash11@gmail.com}{spravash11@gmail.com}}

\author[0000-0002-2808-0888]{Pham Ngoc Diep}
\affiliation{Department of Astrophysics, Vietnam National Space Center, Vietnam Academy of Science and Technology, 18 Hoang Quoc Viet, Hanoi, Vietnam}
\affiliation{Graduate University of Science and Technology, Vietnam Academy of Science and Technology, 18 Hoang Quoc Viet, Hanoi, Vietnam}

\author[0000-0003-2017-0982]{Thiem Hoang}
\affiliation{Korea Astronomy and Space Science Institute, 776 Daedeokdae-ro, Yuseong-gu, Daejeon 34055, Republic of Korea}
\affiliation{University of Science and Technology, Korea, 217 Gajeong-ro, Yuseong-gu, Daejeon 34113, Republic of Korea}

\author[0000-0002-5913-5554]{Nguyen Bich Ngoc}
\affiliation{Department of Astrophysics, Vietnam National Space Center, Vietnam Academy of Science and Technology, 18 Hoang Quoc Viet, Hanoi, Vietnam}
\affiliation{Graduate University of Science and Technology, Vietnam Academy of Science and Technology, 18 Hoang Quoc Viet, Hanoi, Vietnam}

\author[0000-0002-6488-8227]{Le Ngoc Tram}
\affiliation{Leiden Observatory, Leiden University, PO Box 9513, 2300 RA Leiden, The Netherlands}

\begin{abstract}
 
 Polarization of starlight and thermal dust emission due to aligned non-spherical grains helps us to trace magnetic field (B-field) morphology in molecular clouds and to study grain alignment mechanisms. In this work, we study grain alignment and disruption mechanisms in a filamentary infrared dark cloud G34.43+0.24 using thermal dust polarization observations from JCMT/POL-2 at 850 $\mu$m. We study in three sub-regions as North harboring MM3 core, Center harboring MM1 and MM2 cores and South having no core. We find the decrease in polarization fraction $P$ with increasing total intensity and gas column density, known as polarization hole. To disentangle the effect of magnetic field tangling on the polarization hole, we estimate the polarization angle dispersion function. We find depolarizations in North and Center regions are due to decrease in net alignment efficiency of grains but in South region, effect of magnetic field tangling is significant to cause depolarization. To test whether RAdiative Torque (RAT) mechanism can reproduce the observational data, we calculate minimum alignment and disruption sizes of grains using RAT theory and our study finds that RAT alignment mechanism can explain the depolarizations in North and Center regions where B-field tangling effect is less important, except for core regions. We find hints of RAdiative Torque Disruption (RAT-D) in the core regions of MM3 in North, MM1 and MM2 in Center. We also find that the high $P$ value of around 8-20$\%$ in the outer regions of the filament can be explained potentially by magnetically enhanced RAT alignment mechanism.
 
\end{abstract}

\keywords{Interstellar dust (836); Interstellar filaments (842); Star forming regions (1565); Interstellar magnetic fields (845)}

\section{\bf{Introduction}} \label{section:introduction}
Dust grains in the interstellar medium have significant influences in various processes including star and planet formation, gas heating, and cooling and they also provide the surface for the formation of different complex molecules (see \citealt{2003ARA&A..41..241D}). The polarization of starlight due to dichroic extinction by non-spherical grains aligned with the interstellar magnetic field was first observed by \cite{1949Sci...109..166H} and \cite{1949ApJ...109..471H}. Such aligned dust grains also re-emit polarized thermal emission at longer wavelengths \citep{1988QJRAS..29..327H}. Dust polarization has been widely used to trace the magnetic fields in different environments, from the diffuse ISM to star-forming molecular clouds, and also to study dust properties like shape, size, composition, etc. (e.g, \citealt{2021ApJ...919...65D}). Magnetic fields are thought to have significant impact on the formation and evolution of molecular clouds and on the regulation of star formation processes (see \citealt{2012ARA&A..50...29C, 2019FrASS...6...15P}).
Dust grains are aligned in such a way that their short axes are parallel and long axes are perpendicular to the ambient magnetic field (e.g., \citealt{2007JQSRT.106..225L, 2015ARA&A..53..501A, 2015psps.book...81L}). For the starlight polarization induced by the dichroic extinction by aligned grains, the observed polarization angle provides the plane-of-sky (POS) component of magnetic fields. However, for the polarization of thermal dust emission, the observed polarization angle needs to be rotated by $90^\circ$ to infer the POS projected magnetic fields (e.g., \citealt{1988QJRAS..29..327H}). 

Recent studies suggest the potential of tracing three-dimensional (3D) magnetic fields using full dust polarization data including the polarization angle and the polarization fraction \citep{2024ApJ...965..183H, 2024arXiv240714896T}. This technique relies on the comparison of the observed dust polarization with the accurate model of the dust polarization predicted using the grain alignment physics and dust properties (grain shape and composition). Therefore, it is essential to study in detail the exact physical mechanisms for grain alignment. Several mechanisms for grain alignment have been introduced but many of them are not able to explain fully the observational results (see \citealt{2007JQSRT.106..225L} for a review). The most acceptable mechanism which can explain observational results in different environments is the Radiative Torque Alignment (RAT-A) first introduced by \cite{1976Ap&SS..43..291D} and numerically demonstrated in \cite{1997ApJ...480..633D}. An analytical RAT theory was later developed by \cite{2007MNRAS.378..910L} and \cite{2008MNRAS.388..117H}. According to the RAT-A theory, non-spherical grains exposed to anisotropic radiation field acquire radiative torques which tend to spin up grains to suprathermal rotation and align them with the ambient magnetic fields \citep{1997ApJ...480..633D, 2007MNRAS.378..910L}. The efficient alignment of grains is achieved when grains rotate suprathermally with the rotation rate exceeding about 3 times of their thermal angular velocity \citep{2008MNRAS.388..117H, 2016ApJ...831..159H}. This RAT-A theory predicts an anti-correlation of the polarization fraction $P$ with local gas density and a correlation of $P$ with the radiation field intensity or dust temperature (see \citealt{2020ApJ...896...44L, 2021ApJ...908..218H}). Observations in various regions found a trend of decreasing polarization fraction with the increase in gas column density or total intensity, termed as polarization hole (see e.g, \citealt{2019FrASS...6...15P, 2021ApJ...908..218H}). Various tests of RAT-A in molecular clouds using starlight polarization support the theory well (see review by \citealt{2015ARA&A..53..501A}).

However, the test of RAT-A mechanism in star-forming regions using thermal dust polarization is only conducted recently \citep{2022FrASS...9.3927T, 2023ApJ...953...66N} thanks to the advance in far-IR and sub-mm/mm polarimetric instruments HAWC+ \citep{2018JAI.....740008H}, POL-2 \citep{2016SPIE.9914E..03F}, ALMA \citep{2009IEEEP..97.1463W}, NIKA-2 \citep{2018A&A...609A.115A}, etc. In the study of dust polarization in different star-forming regions, the polarization hole may be caused not only by decrease in grain alignment in denser regions as predicted by RAT-A but also by magnetic field fluctuations along the line-of-sight. Turbulences in the molecular clouds are the main cause of magnetic field fluctuations \citep{1989ApJ...346..728J, 1992ApJ...389..602J, 2008ApJ...679..537F}.

Based on the advance in RAT-A theory, \cite{2019NatAs...3..766H} introduced a new mechanism termed as RAdiative Torque Disruption (RAT-D). From this RAT-D mechanism, when large grains are irradiated by radiation of very high intensity they rotate very fast suprathermally as induced by radiative torques and when the centrifugal stress is greater than the tensile strength of the grains, the large grains could be disrupted into smaller fragments. This RAT-D effect can decrease the polarization fraction when the dust temperature is sufficiently large, because of the depletion of large grains \citep{2020ApJ...896...44L} and as RAT-A becomes less effective to smaller grains. Many studies reported that polarization fraction $P$ first increases with dust temperature $T_\mathrm{d}$ and when $T_\mathrm{d}$ becomes larger, after some $T_\mathrm{d}$ value $P$ decreases (e.g, \citealt{2020A&A...641A..12P, 2019ApJ...882..113S}). This decreasing of $P$ with $T_\mathrm{d}$ poses a challenge to the popular RAT-A theory and could be well explained by the RAT-D mechanism. A comprehensive testing for the RAT paradigm (RAT-A and RAT-D) in molecular clouds and interstellar filaments is first studied in
\cite{2021ApJ...906..115T}, \cite{2024ApJ...974..118N} and \cite{2024arXiv240317088T}. In this paper, we aim to test the RAT paradigm for an infrared dark cloud (IRDC) that contains embedded active star-forming protostellar cores. The decrease in $P$ at larger $T_\mathrm{d}$ values can be expected near the high radiation field of protostellar cores in such IRDC due to RAT-D effect or magnetic field tangling along the line-of-sight. 

Infrared Dark Clouds (IRDCs) are usually associated with the formation of mostly high-mass stars and star clusters \citep{2006ApJ...641..389R} and they contain mostly cold and dense molecular gas. For the study of grain alignment mechanisms, the physical parameters like gas column density or volume density and dust temperature or equivalently the radiation field strength are crucial. A significant variation of these parameters can favor for the grain alignment study. The filamentary IRDCs show significant variations in these parameters across the filaments from the outer regions towards the inner regions which makes them ideal regions for studying grain alignment mechanisms. In this work, we study the grain alignment mechanism in the filamentary structured IRDC G34.43+0.24 (hereafter G34) located at a distance of nearly 3.7 kpc \citep{2012ApJ...756...60S, 2014ApJ...791..108F} and is an active high-mass star-forming cloud \citep{1998A&A...336..339M, 2011ApJ...741..120R, 2018ApJ...857...35S}. This filament consists of many cores MM1 to MM9 which are likely at different stages of evolution \citep{2011ApJ...743..196C}. The most evolved core MM2 \citep{2006ApJ...641..389R} is associated with the ultra-compact HII region IRAS 18507+0121 having spectral type B0.5 \citep{1998A&A...336..339M, 2004ApJ...602..850S, 2007ApJ...669..464S}. MM1 is the brightest millimeter core with hints of high-mass core \citep{2008ApJ...689.1141R} and MM3 hosts a hot-corino \citep{2014ApJ...794L..10Y, 2015ApJ...803...70S}. We want to study the grain alignment mechanism over all the filament which has significant variations in gas density and dust temperature by categorizing three sub-regions as North, Center and South. The central region containing cores MM1 and MM2 is very bright and highly dense as compared to other regions of the filament. The North and Center regions harbor cores but the South region has no cores. This filament is ideal for testing the RAT-A and RAT-D because some regions contain embedded sources and some do not. The rest of the paper is organized as follows: Section \ref{sec:Data} gives the data acquisition details, Section \ref{section:Results and Analysis} for the analysis and results, Section \ref{section:Discussions} presents the discussions of the results and Section \ref{Section:Conclusions} gives the summary of the work.  
\begin{figure*}
    \centering
    \begin{tabular}{cc}
        \includegraphics[scale=0.56]{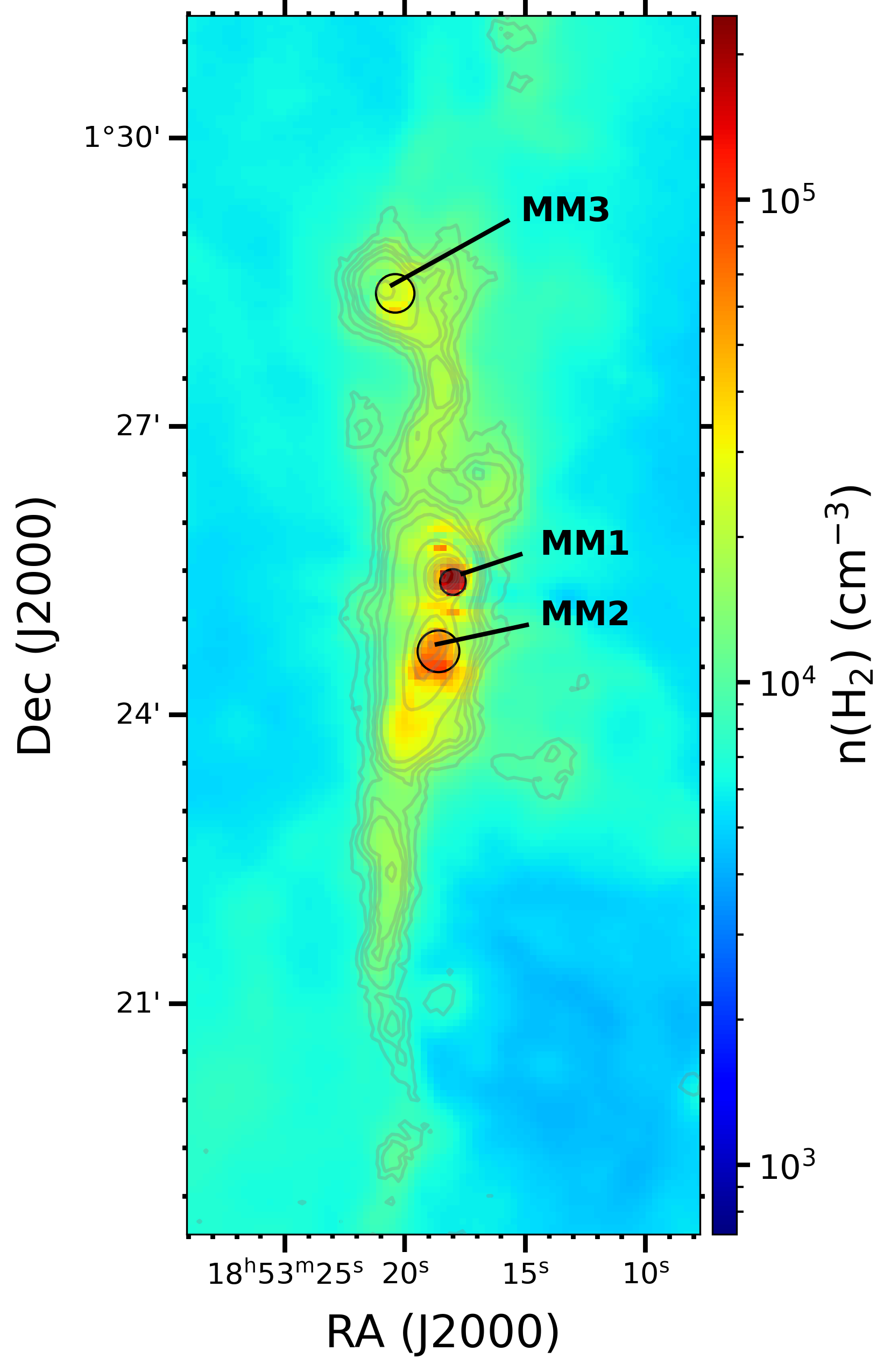} & 
        \hspace{20pt}
        \includegraphics[scale=0.56]{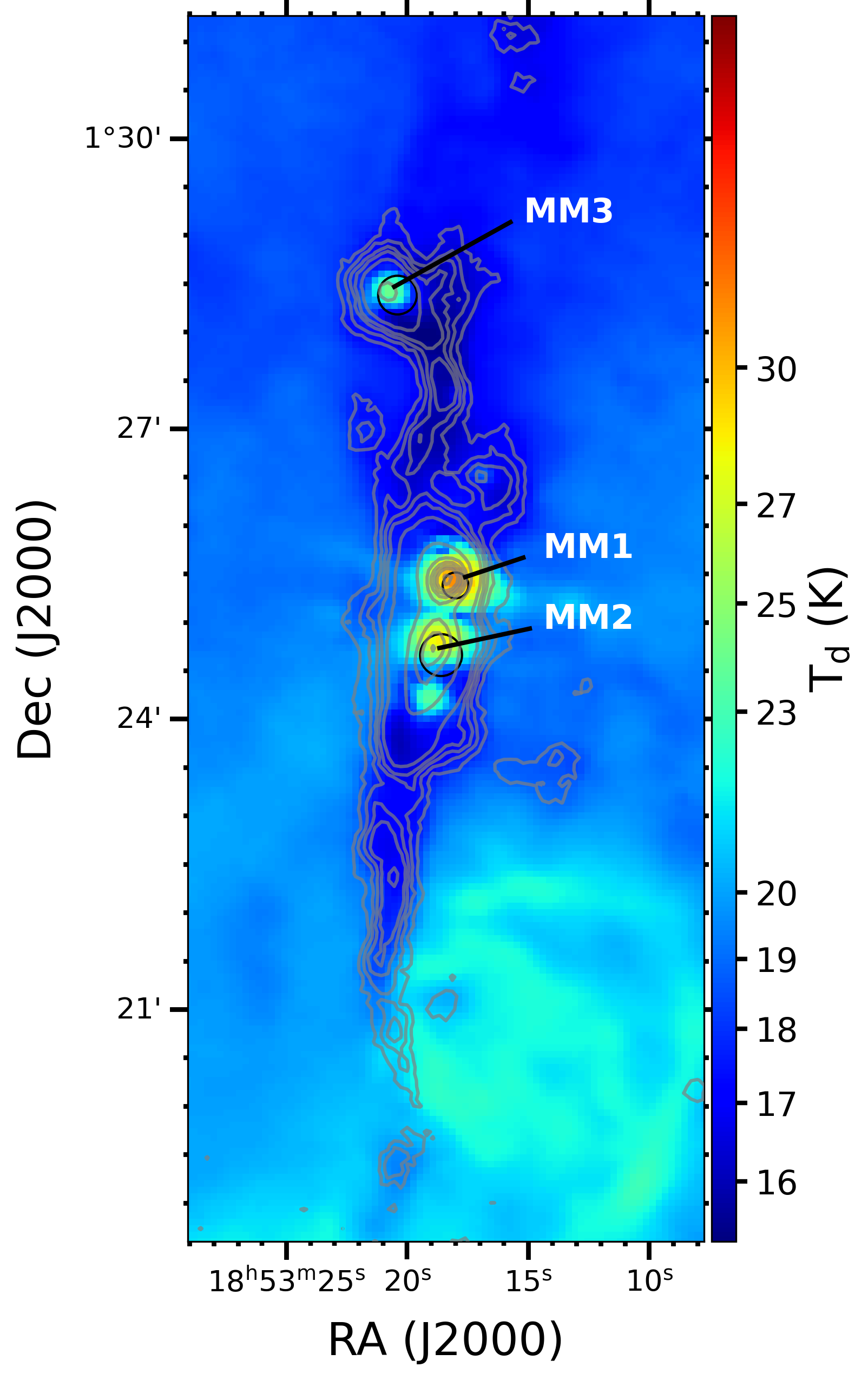} 
    \end{tabular}
    \caption{Maps of $\mathrm{H_{2}}$ volume density (left) and dust temperature (right) for G34 filament. The locations of the cores MM1, MM2 and MM3 as identified by \cite{2006ApJ...641..389R} are indicated with black circles.}
    \label{Fig:vd_Td_map}
\end{figure*}

\begin{figure*}
    \centering
        \includegraphics[scale=0.6]{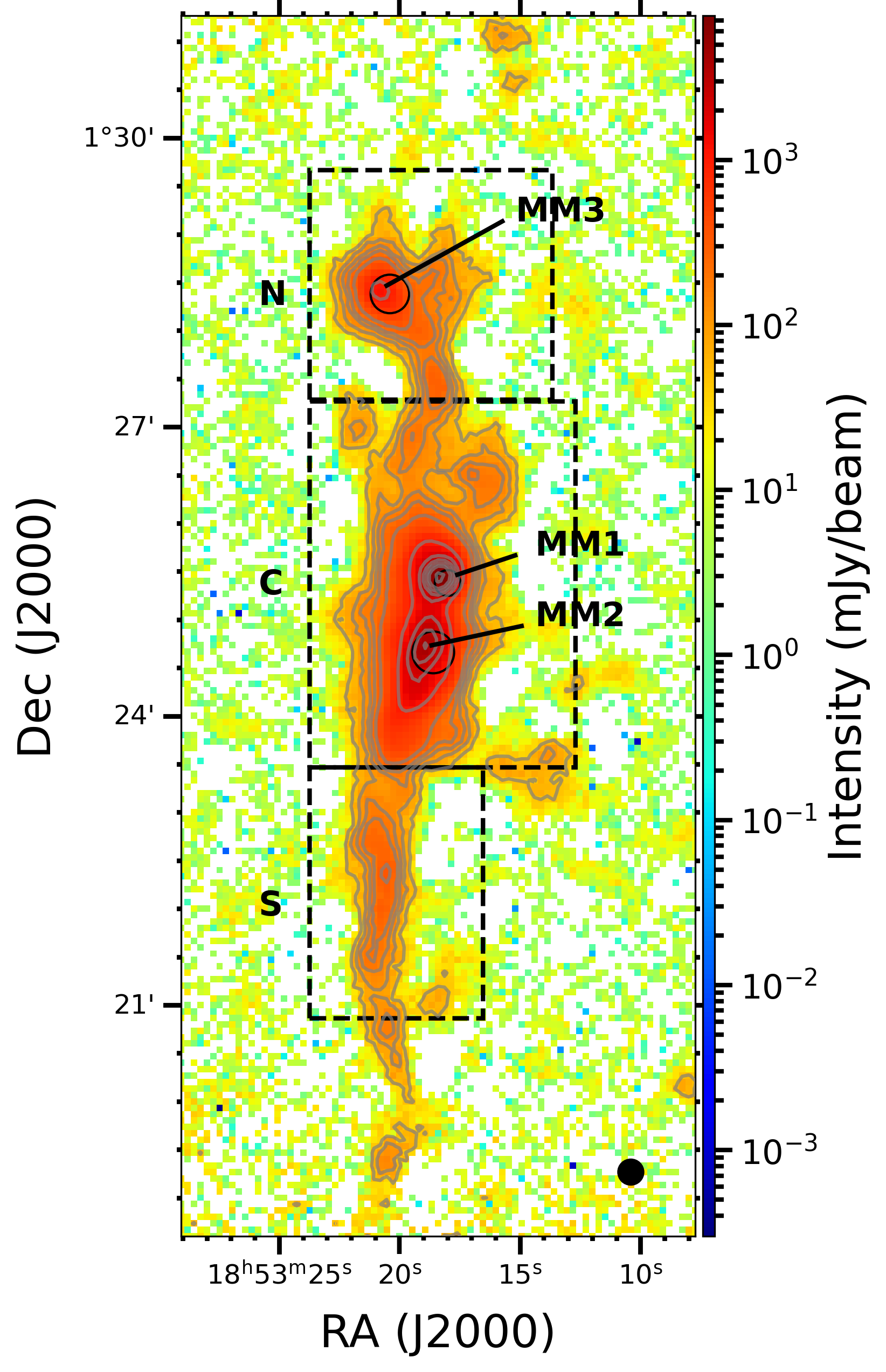} 
    \caption{Total emission intensity ($I$) map of the G34 filament observed by JCMT/POL-2 at 850$\mu$m with overlaid contours drawn at $I$ values of 50, 100, 150, 200, 300, 1000, 2000, 3000, 4000, 5000, 7000 and 8000 mJy/beam. The positions of three dense millimeter cores MM1, MM2 in the Center and MM3 in the North regions are indicated with black circles. Also, the beam size, $14''.1$ of JCMT/POL-2 at 850$\mu$m is indicated with a solid black circle.}
    \label{Figure:Intensity_map}
\end{figure*}

\section{\bf{Data Acquisition}} \label{sec:Data}
\subsection{Archival polarization data}
In this work, we use the archival dust polarization data observed by the POL-2 instrument mounted on the James Clerk Maxwell Telescope (JCMT) at 850 $\mu$m. The data is taken from \cite{2019ApJ...883...95S}. The observations were conducted in 2018 August in band-2 weather conditions using the POL-2 daisy map mode of JCMT. The Full-Width Half Maximum (FWHM) of JCMT/POL-2 at 850 $\mu$m is $14''.1$. The data were reduced using the STARLINK/SMURF package $pol2map$. The polarization data are debiased using the mean of variances of $Q$ and $U$ to remove statistical bias in the regions of low signal-to-noise ratio. The debiased polarization fraction and the uncertainty are obtained using Equation 1 for debiased polarization fraction and Equation 2 for the uncertainty in the debiased polarization fraction as given in \cite{2019ApJ...883...95S}. The selection criteria on the total intensity is kept on $I/\sigma_I > 10$. The quality of the data was checked using different signal-to-noise ratio, $S/N > 2$ $(PI/\sigma_{PI} > 2)$ and $S/N > 3$ $(PI/\sigma_{PI} > 3)$ where $PI$ is the polarization intensity and $\sigma_{PI}$ is its uncertainty. In panel (c) and (d) of Figure 2 in \cite{2019ApJ...883...95S}, the histogram distributions of magnetic field position angles ($PA$) and polarization fraction ($P$) show very similar trends for both $S/N > 2$ and $S/N > 3$ which makes to consider the $2 < S/N < 3$ data for further analysis. For our analysis, we use 211 polarization vectors satisfying signal-to-noise ratio $S/N > 2$ same as used in the study of magnetic field morphology and strength in \cite{2019ApJ...883...95S}. For the details on data acquisition, reduction, validation, and selection, please refer to Section 2 of \cite{2019ApJ...883...95S}.

\subsection{Maps of $H_2$ column densities, volume densities and dust temperatures} \label{sec:column_density}
We use the $\mathrm{H_2}$ column density and the dust temperature maps of G34 filament as estimated in  \cite{2019ApJ...883...95S} using archival $Herschel$ PACS/SPIRE 70, 160, 250, 350, 500 $\mu$m and JCMT 850 $\mu$m data fitted with a modified blackbody function.

For calculating the $\mathrm{H_2}$ volume densities, we use the column density map. \cite{2024ApJ...960...76P} estimated the width of G34 filament to be $\approx$ 0.6 pc using $RadFil$ algorithm \citep{2018ApJ...864..152Z}. For the cores MM1, MM2 in the Center region and MM3 in the North region, the angular FWHM diameters are estimated to be 16$''$, 26$''$ and 24$''$ that corresponds to physical FWHM diameters of around 0.19 pc, 0.42 pc and 0.38 pc, respectively \citep{2006ApJ...641..389R}. We assume a cylindrical geometry of the overall filament so that the depth of the filament is equal to its width. However, we assume the cores to have spherical geometries so that their widths or diameters are equal to their depths. We use 0.6 pc as the depth value for the overall filament except the core regions and values of 0.19 pc for MM1 core, 0.42 pc for MM2 core and 0.38 pc for MM3 core to derive the volume densities of the filament. The volume densities $n(\mathrm{H_2})$ are calculated as follows

\begin{equation}
{
n(\mathrm{H_2}) = \frac{N(\mathrm{H_2})}{d},
}
\end{equation}
where $d$ is the depth of the filament. Also, it is noted that the estimation of width or depth of filaments is biased based on the angular resolution of the observations \citep{2022A&A...657L..13P}. The maps of $n(\mathrm{H_2})$ and $T_\mathrm{d}$ are shown in Figure \ref{Fig:vd_Td_map}. The volume densities are found to be in the range from $10^4$ to $10^5$ $\mathrm{cm^{-3}}$. The Center region containing the MM1 and MM2 cores has very high gas volume densities of the order of $10^5$ $\mathrm{cm^{-3}}$ and the region of MM3 core has volume density of the order of $10^4$ $\mathrm{cm^{-3}}$. The dust temperature decreases from the outer regions towards the inner regions of the filament except for the regions of MM1, MM2 and MM3 cores which show high temperatures up to around 31 K. Therefore, the dust grains in the regions inside the filament except the MM1, MM2 and MM3 cores are heated only by the diffused interstellar radiation field (ISRF) whereas those in the regions of MM1, MM2 and MM3 cores are heated by both the ISRF and the internal radiation field from these protostellar cores.

\section{\bf{Analysis and Results}} \label{section:Results and Analysis}
\subsection{Data Analyses}
\subsubsection{Total emission intensity map} \label{sec:Intensity_map}
The map of the total emission intensity $I$ of G34 filament observed by the JCMT/POL-2 instrument at 850$\mu$m is shown in Figure \ref{Figure:Intensity_map}. We mark three sub-regions over the filament with three black dotted rectangles as N, C and S to denote the North, Center and South regions of the filament respectively in the same way as in \cite{2019ApJ...883...95S}. The positions of the millimeter cores MM1, MM2 in the Center and MM3 in the North are also indicated in the figure. The regions of MM1 and MM2 in the Center region show a very high intensity. The MM3 region in the North also shows high intensity but is comparatively lower than the MM1 and MM2 regions. The intensity increases from the outer regions to the inner regions of the filament. 

\subsubsection{Polarization fraction map} \label{sec:P_map}
\begin{figure*}
    \centering
    \begin{tabular}{cc}
        \includegraphics[scale=0.49]{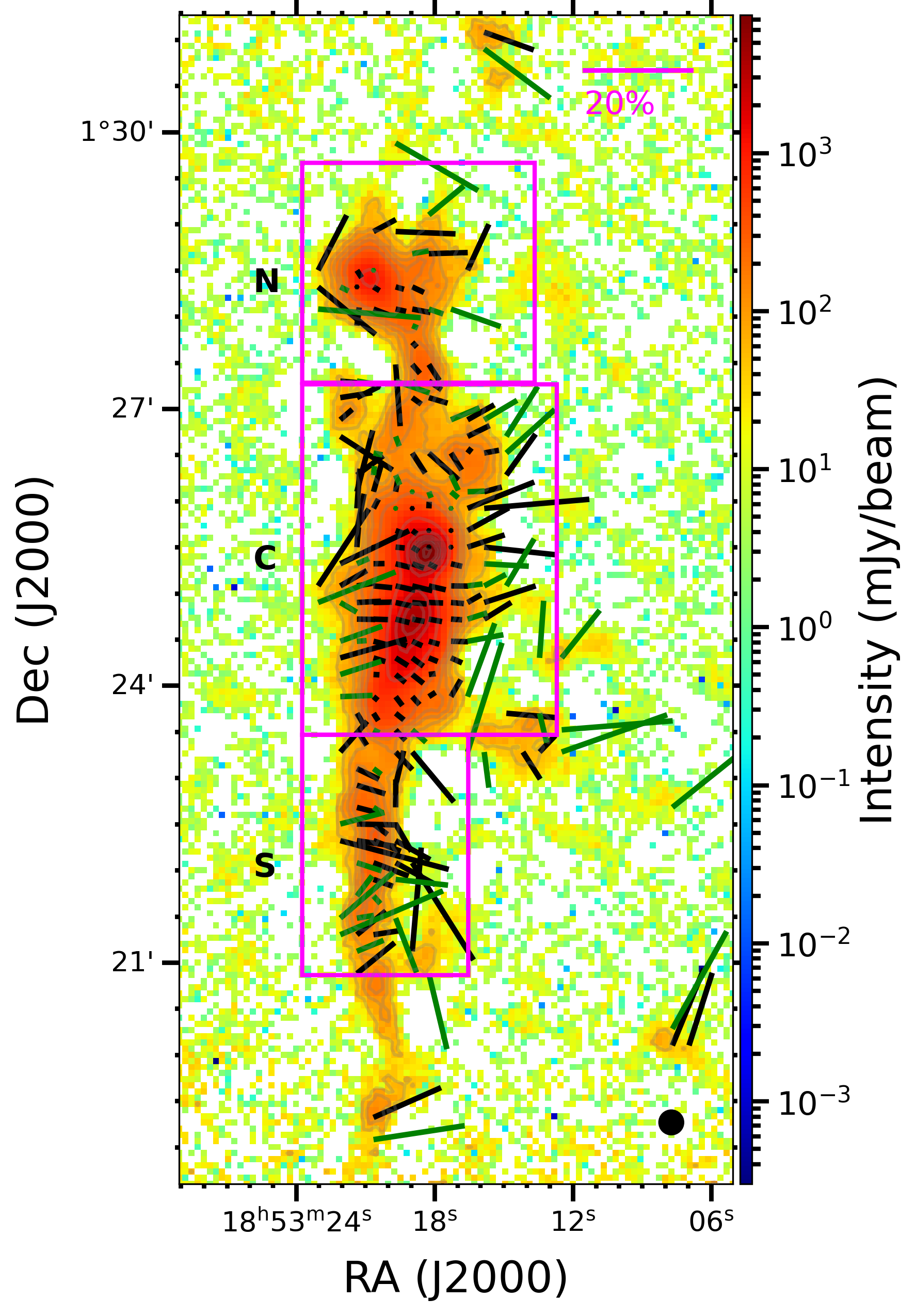} & 
        \hspace{5pt}
        \includegraphics[scale=0.56]{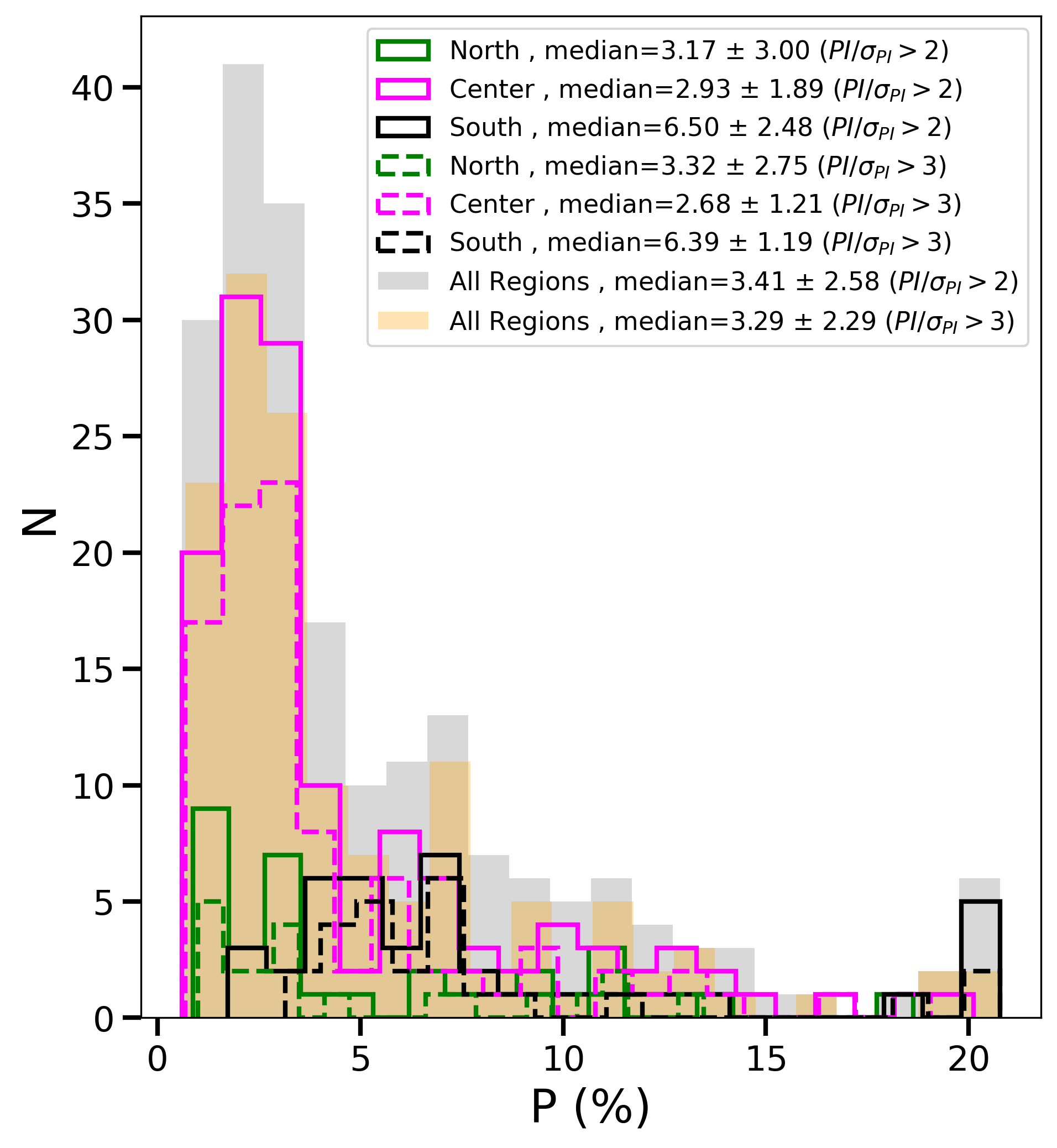} 
    \end{tabular}
    \caption{Map of distribution of polarization vectors over all the G34 filament with the black vectors corresponding to $PI/\sigma_{PI} > 3$ and green vectors for $2 < PI/\sigma_{PI} < 3$ (left panel) and histograms of polarization fraction $P$ for the North (green), Center (magenta), South (black) regions with solid lines for $PI/\sigma_{PI} > 2$ and dashed lines for $PI/\sigma_{PI} > 3$, the gray-filled for all regions with $PI/\sigma_{PI} > 2$ and orange-filled for all regions with $PI/\sigma_{PI} > 3$ (right panel). The lengths of the vectors are proportional to $P$ and their orientations determine the magnetic field orientations. A reference scale length of 20$\%$ for the polarization fraction is indicated. The contours are the same as in Figure \ref{Figure:Intensity_map}.}
    \label{Fig:P_map_Histogram_P}
\end{figure*}

The map of the distribution of the polarization vectors observed by JCMT/POL-2 at 850 $\mu$m over the G34 filament with $PI/\sigma_{PI}$ $> 3$ in black colors and $2 < PI/\sigma_{PI} < 3$ in green colors is given in Figure \ref{Fig:P_map_Histogram_P} (left panel). The lengths of the vectors are proportional to the polarization fraction $P$ and their orientations determine the magnetic field orientations. We find that $P$ becomes smaller and nearly constant in the inner regions where the dust emission intensities are higher and becomes larger in the outer diffused regions of the filament where the intensities are lower. We find high polarization fraction values of 8-20\% in the outer regions having very small Stokes $I$ values. Some outer regions have very high $P$ values going up to around 20 $\pm$ 6\% (typical uncertainty). This feature is also observed in several studies using POL-2 and HAWC+ observations (e.g, \citealt{2018ApJ...859....4K, 2018ApJ...861...65S, 2019ApJ...877...88C, 2019ApJ...880...27P, 2019ApJ...876...42W, 2021A&A...647A..78A, 2023ApJ...953...66N}) and in the protostellar envelopes using ALMA observations (e.g, \citealt{2017ApJ...847...92H, 2019ApJ...879...25K}). We marked three sub-regions as N, C and S with three rectangles in the figure indicating North, Center and South regions of the filament respectively, in the same way as we define in Figure \ref{Figure:Intensity_map} to study the grain alignment mechanisms in each of these sub-regions. 

The histogram distributions of the polarization fraction $P$ for each sub-region are given in Figure \ref{Fig:P_map_Histogram_P} (lower panel). The solid and the dashed green lines are for the North, magenta lines for the Center and black lines for the South regions with $S/N > 2$ and $S/N > 3$, respectively. The filled gray color is for all the regions with $S/N > 2$ and filled orange color for all the regions with $S/N > 3$. The median polarization fractions for the North, Center and South regions with $S/N > 2$ are $3.17 \pm 3.00$\%, $2.93 \pm 1.89$\%, $6.50 \pm 2.48$\% and for $S/N > 3$ are $3.32 \pm 2.75$\%, $2.68 \pm 1.21$\%, $6.39 \pm 1.19$\%, respectively. The median polarization fractions for all the regions with $S/N > 2$ and $S/N > 3$ are $3.41 \pm 2.58$\% and $3.29 \pm 2.29$\%, respectively. The median values for both $S/N > 2$ and $S/N > 3$ are nearly similar.

\begin{figure}
    \centering
    \includegraphics[width=0.46\textwidth]{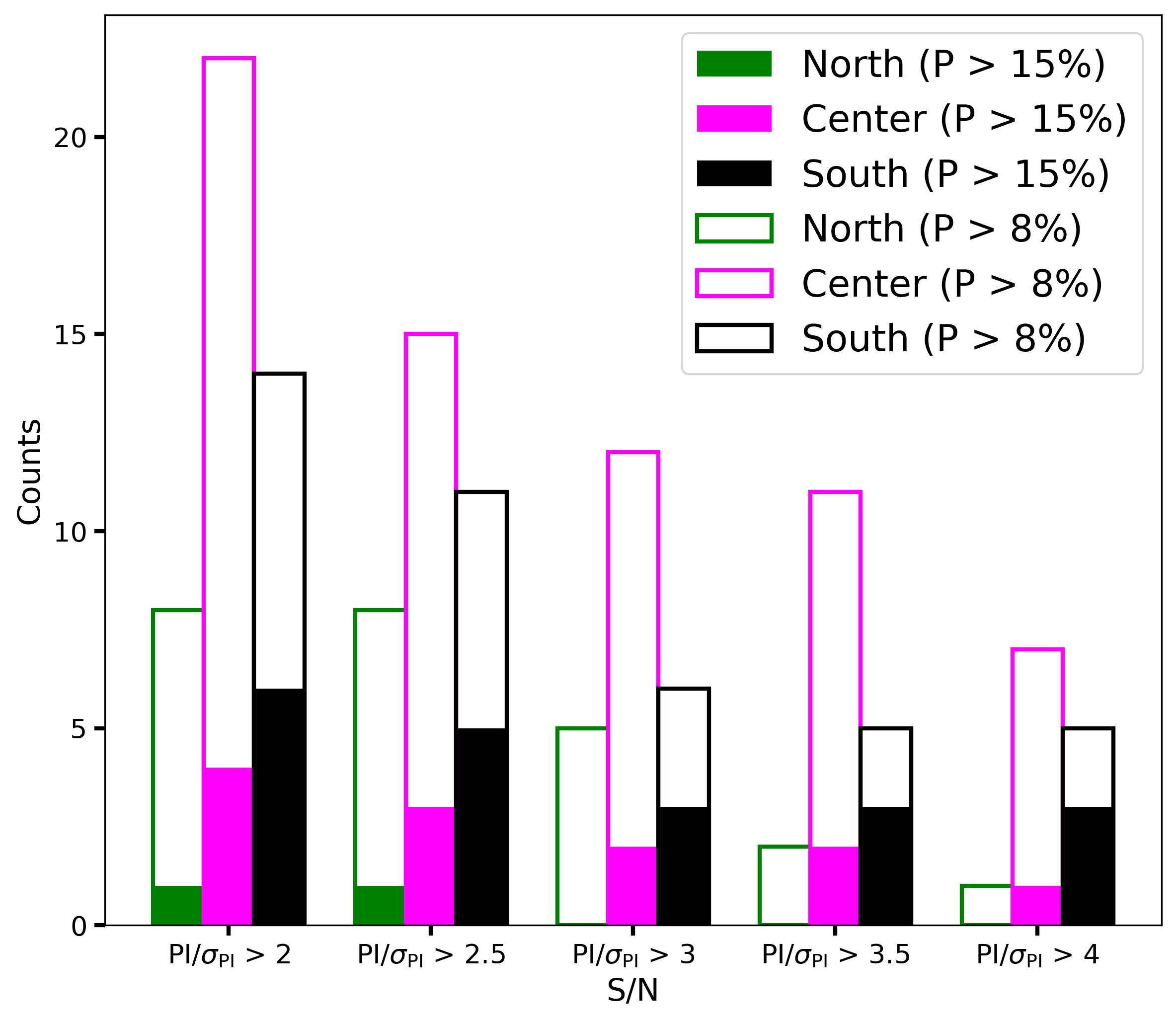}
    \caption{Bar graph showing the counts of $P > 15$\% and $P > 8$\% in each region for different signal-to-noise ratios.}
    \label{Fig:Bar_diagram}
\end{figure}

Figure \ref{Fig:Bar_diagram} shows the counts of $P > 15$\% and $P > 8$\% in the North, Center and the South regions for different S/N of $PI/\sigma_{PI} >$ 2, 2.5, 3, 3.5 and 4 just to see the trend of high polarization values with increasing S/N. There are some pixels with high $P$ values of around $20 \pm 6\%$ in the outer regions of the filament but more number in the outer parts of the South region. The Center and the North regions have pixels in the outer regions with high $P$ values of more than 8\%. The Center region seems to have overall $P$ values lower than the North and South regions.

\begin{figure*}
    \centering
    \includegraphics[width=0.85\textwidth]{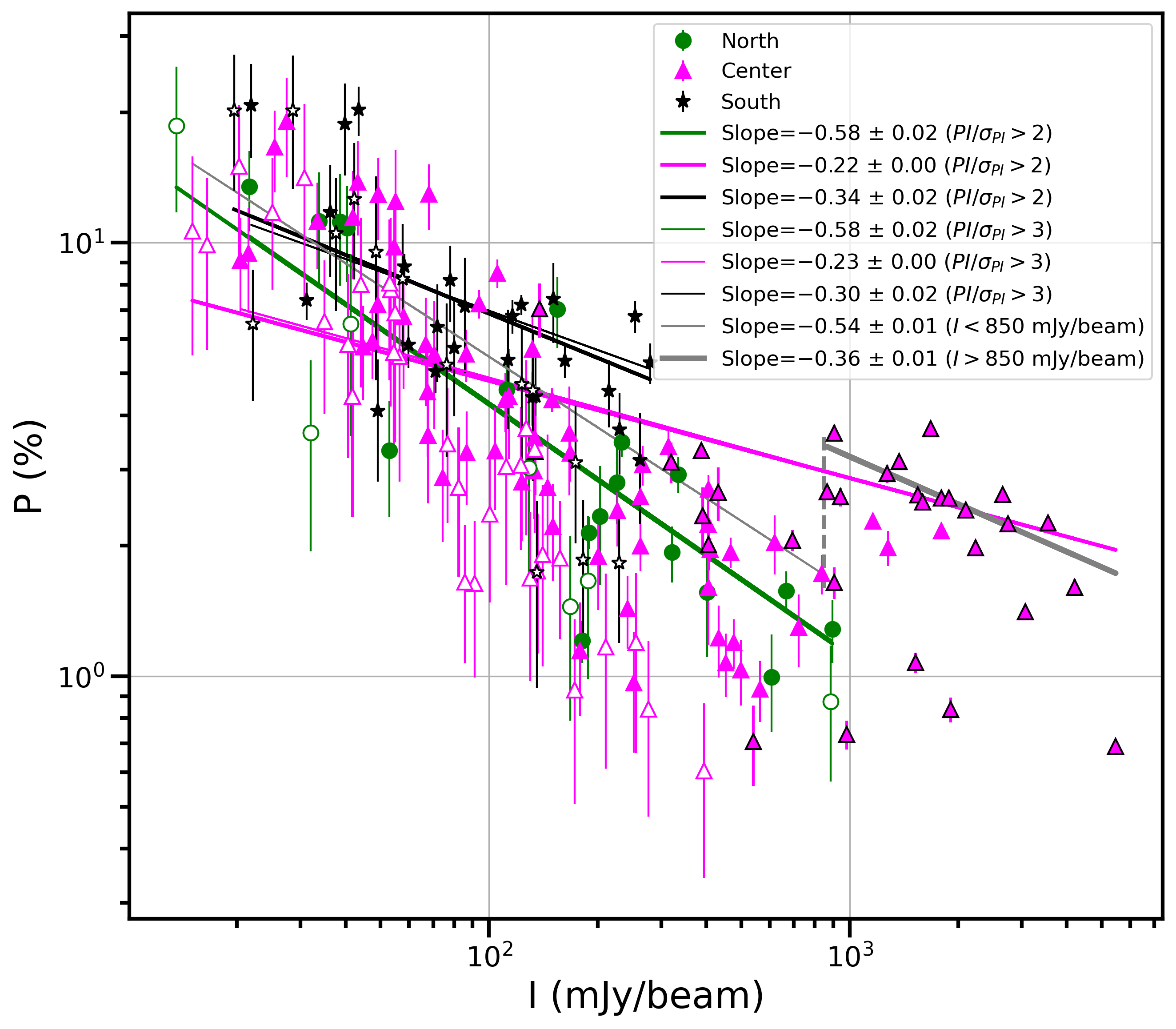}
    \caption{Variations of polarization fractions $P$ with the total intensities $I$ for the North, Center and South regions. The green, magenta and black data points are for the North, Center and the South regions with $PI/\sigma_{PI} > 3$ and the white facecolor data points are for $2 < PI/\sigma_{PI} < 3$. The data points in the Center region with $T_\mathrm{d} > 19.8$ K are shown with black edgecolors. The solid lines are the weighted best-fit lines for each region with same color notation. The thin and thick gray solid lines are fitted for $I < 850$ mJy/beam and $I > 850$ mJy/beam for the Center region with the gray vertical dashed line to visualize the increase in $P$ at around 850 mJy/beam.} $P$ is found to decrease in each of the regions overall.
    \label{Figure:P_I}
\end{figure*}

\subsubsection{Polarization fraction as a function of total intensities} \label{sec:P_I}
In Figure \ref{Figure:P_I}, the polarization fraction $P$ is very high up to nearly 8-20$\%$ in the outer regions having total intensities $I$ less than 60 mJy/beam but decreases as $I$ increases towards the filament's spine. We check the variations of $P$ with $I$ for both $S/N > 2$ and $S/N > 3$. The data for $2 < S/N < 3$ are shown with white facecolors keeping the edgecolors in same notation as we use to denote North, Center and South regions. Also, in the Center region we identify those data points corresponding to higher dust temperature values of greater than 19.8 K with black edgecolors (see Section \ref{section:P_NH2_Td} for this particular dust temperature value). The variations of $P$ with $I$ at the regions of North, Center and South are fitted with weighted power-law fits of the form $P=k_1I^{a_1}$, where $k_1$ is a constant and $a_1$ gives the slope values of the fits as $-0.58 \pm 0.02$, $-0.22 \pm 0.00$ and $-0.34 \pm 0.02$ for $S/N > 2$ and $-0.58 \pm 0.02$, $-0.23 \pm 0.00$ and $-0.30 \pm 0.02$ for $S/N > 3$ respectively with the uncertainties on the values of $a_1$ resulting from the best fits. We find that the weighted fits in all the regions for both $S/N > 2$ and $S/N > 3$ are almost similar. Also, since the fit is weighted fit, it mostly favors the higher $S/N$ data with less favor on the lower $S/N$ data. For further analysis in this work, we use $S/N > 2$ data with weighted fits. The overall slope in the Center region is shallower than the North and the South regions. For the Center region, we also fit weighted power-law fits for $I < 850$ mJy/beam and $I \geq 850$ mJy/beam with thin and thick gray solid lines giving slope values of $-0.54 \pm 0.01$ and $-0.36 \pm 0.01$, respectively. At this $I=850$ mJy/beam, there is an increase in $P$ shown with a gray dashed line and then start decreasing upto a $P$ value of around 0.6$\%$. These may be explained in such a way that the decrease in $P$ upto 850 mJy/beam may be associated with the regions outside the cores and may be due to decrease in RAT alignment efficiency of grains or magnetic field tangling along the line of sight.
\begin{figure*}
    \centering
    \includegraphics[width=0.8\textwidth]{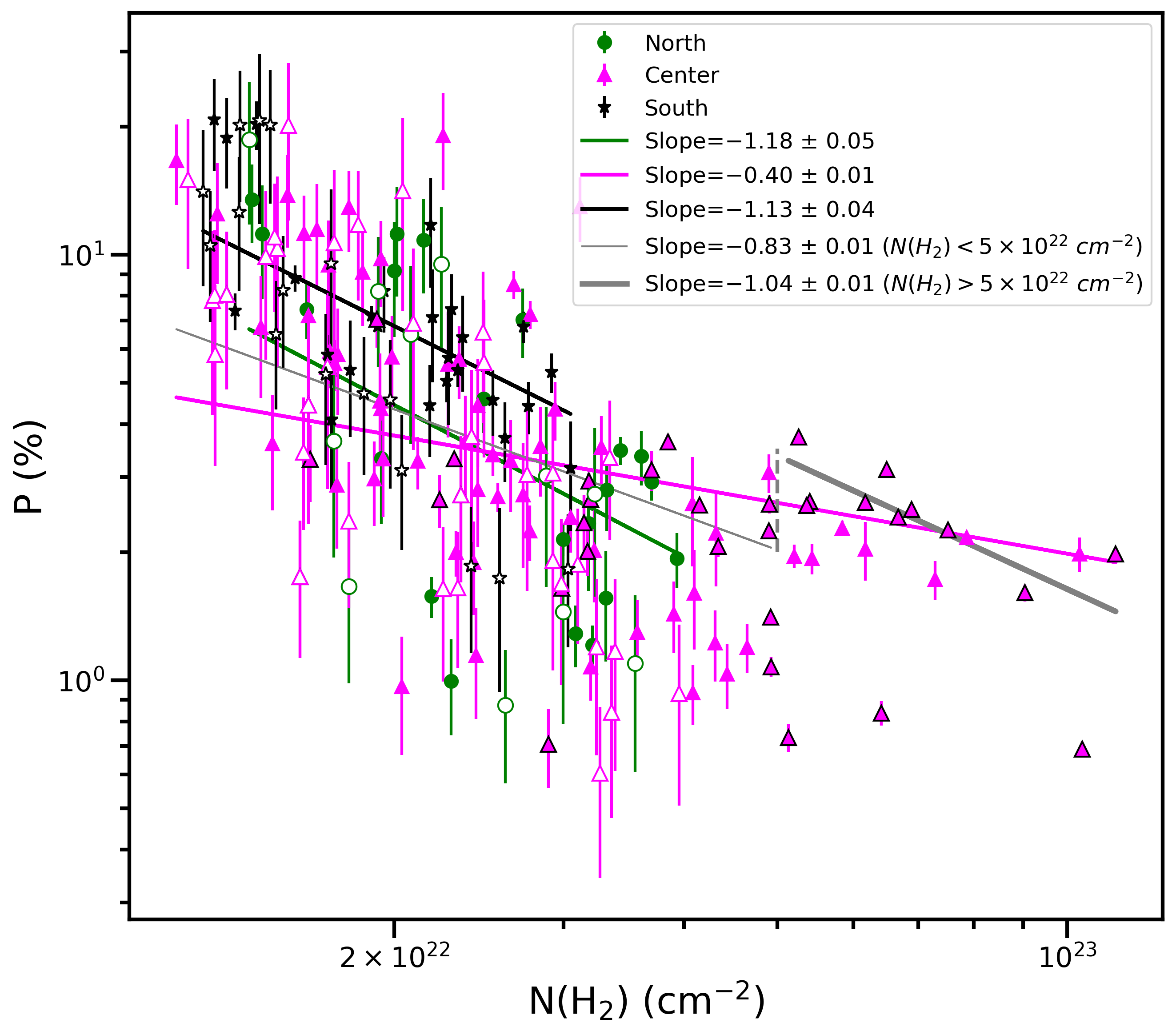}
    \caption{Variations of polarization fractions $P$ with the gas column densities $N(\mathrm{H_2})$ for the North, Center and South regions. The thin and thick gray solid lines are the best-fit lines for $N(\mathrm{H_2}) < 5 \times 10^{22}$ $\mathrm{cm^{-2}}$ and $N(\mathrm{H_2}) > 5 \times 10^{22}$ $\mathrm{cm^{-2}}$. The gray vertical dashed line is drawn to visualize the increase in $P$ at $N(\mathrm{H_2}) = 5 \times 10^{22}$ $\mathrm{cm^{-2}}$. The other notations remain the same as in Figure \ref{Figure:P_I}. $P$ is found to decrease with $N(\mathrm{H_2})$ in each of the regions overall like $P$ decreasing with $I$ in Figure \ref{Figure:P_I}.}
    \label{Figure:P_NH2}
\end{figure*}
The increase in $P$ at around 850 mJy/beam may be associated with the grains found in the regions approaching towards the central cores where the attenuated radiation field from the embedded source favors for the alignment of the grains. The decrease in $P$ after 850 mJy/beam may be associated with the grains in the regions very near to the central protostar where the internal radiation field strength is very high and may disrupt the large grains in these dense and hot regions (see Figure 12 in \citealt{2021ApJ...908..218H}). The decrease in $P$ may also be due to magnetic field tangling along the line of sight.

The parameter $a_1$ is a very significant parameter that gives information on the grain alignment efficiency variation and magnetic field tangling along the line-of-sight with its values in molecular clouds generally ranging from 0 to 1 (0 for perfect alignment, 1 for no alignment and in between 0 and 1 for some degree of alignment). The most significant source of magnetic field tangling is the turbulence in the cloud \citep{1989ApJ...346..728J, 1992ApJ...389..602J, 2008ApJ...679..537F}.

\subsubsection{Polarization fraction as a function of Column densities and Dust temperatures} \label{section:P_NH2_Td}
\begin{figure*}
    \centering
    \includegraphics[width=0.8\textwidth]{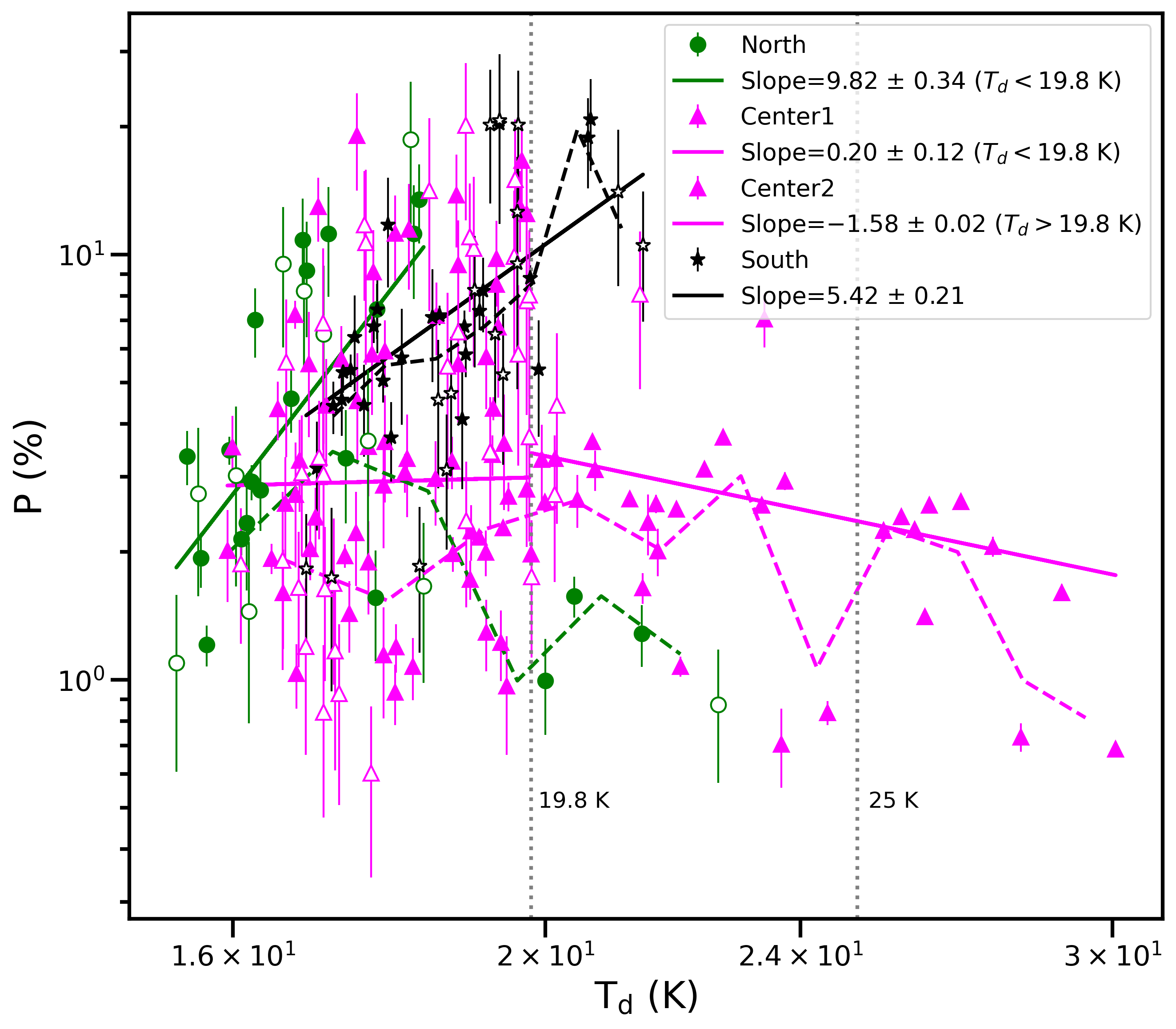}
    \caption{Variations of polarization fractions $P$ with dust temperatures $T_\mathrm{d}$ in the North, Center and South regions. The dashed lines are the running means and the solid lines are the weighted best power-law fits of the respective regions. For the North region, the fit is done only for the data up to 19.8 K. A vertical dotted line is drawn at 19.8 K to mark the observed decrease in $P$ in the North and the Center regions. The Center regions show more significant decrease in $P$ after around 25 K shown with another dotted line at 25 K.}
    \label{Figure:P_Td}
\end{figure*}

Local conditions such as the strength of the radiation field (or equivalently the dust temperature) and the gas densities play important roles towards the variation in the efficiency of grain alignment in accordance with the RAT theory \citep{2021ApJ...908..218H}. Therefore, we study the variations of the observed polarization fraction as a function of column density $N(\mathrm{H_2})$ (Figure \ref{Figure:P_NH2}) and dust temperatures $T_\mathrm{d}$ (Figure \ref{Figure:P_Td}), to find the effects of these local conditions of G34 on the polarization fraction.

We find an anticorrelation in the $P-N(\mathrm{H_2})$ plot describable by a weighted power-law fit of the form $P=k_2N(\mathrm{H_2})^{a_2}$, where $k_2$ is a constant and $a_2$ gives the slope with values of $-1.18 \pm 0.05$ for North, $-0.40 \pm 0.01$ for Center and $-1.13 \pm 0.04$ for South regions. The overall slope $a_2$ for the Center region is much shallower than the North and South regions just like we find in the $P-I$ relation in Figure \ref{Figure:P_I}.
We also fit weighted power-law fits in the Center region for $N(\mathrm{H_2}) < 5 \times 10^{22}$ $\mathrm{cm^{-2}}$ and $N(\mathrm{H_2}) \geq 5 \times 10^{22}$ $\mathrm{cm^{-2}}$ with thin and thick gray solid lines giving the slope values of $-0.83 \pm 0.01$ and $-1.04 \pm 0.01$, respectively. We find similar trends with $P-I$ plot in Figure 4. The steep slope in the Center region upto $5 \times 10^{22}$ $\mathrm{cm^{-2}}$ may be associated with the regions outside the cores and after this value may be with the core regions.

In the $P-T_\mathrm{d}$ plot as shown in Figure \ref{Figure:P_Td}, we find that the polarization fraction $P$ increases with increasing dust temperature in the North up to around 19.8 K and then decreases. For the South region, $P$ only increases with $T_\mathrm{d}$. For the Center region, there is more spread in $P$ upto $T_\mathrm{d}=19.8$ K. However, $P$ increases overall up to around 19.8 K and then decreases with increasing dust temperature with $T_\mathrm{d}=19.8$ K acting as the transition temperature. However, the polarization fraction decreases more significantly with less spread after around 25 K in the Center region. The variations of $P$ with $T_\mathrm{d}$ is fitted with weighted power-law fits of the form $P=k_3T_\mathrm{d}^{a_3}$, where $k_3$ is a constant and $a_3$ gives the slope with values of $9.82 \pm 0.34$ for North with $T_\mathrm{d} < 19.8$ K, $5.42 \pm 0.21$ for South, $0.20 \pm 0.12$ for Center with $T_\mathrm{d} < 19.8$ K and $-1.58 \pm 0.02$ with $T_\mathrm{d} > 19.8$ K. For the Center region, the P values after $I=850$ mJy/beam are correlated with those $P$ values after $T_\mathrm{d}=19.8$ K indicated with black edgecolors in Figure \ref{Figure:P_I}. Also, $P$ does not drop rapidly initially but as the dust temperature becomes larger around $T_\mathrm{d} > 25$ K the value of $P$ significantly decreases. Hence, the regions in the Center region after 19.8 K is associated with the cores. The increase in polarization fraction with increasing dust temperature is in agreement with the prediction of RAT-A theory (\citealt{2020ApJ...896...44L}). However, many observational studies in dense molecular clouds show that the polarization fraction does not always increase with dust temperature. For example, in the study of molecular cloud regions of Aquila Rift, Cham-Musca, Orion and Ophiuchus in the Gould Belt cloud with the polarization degree measured by the $Planck$ \citep{2020A&A...641A..12P} it was found that $P$ decreases for $T_\mathrm{d} > 19$ K. Also, in the study of the molecular cloud Ophiucus A \citep{2019ApJ...882..113S, 2021ApJ...906..115T}, $P$ decreases for $T_\mathrm{d} > 25-32$ K. This feature is found in the North and Center regions of G34 filament containing dense and hot cores MM3 (North), MM1 and MM2 (Center).  

When the radiation strength becomes very strong, large grains can be disrupted into smaller fragments because of RAT-D effect \citep{2019NatAs...3..766H}. This effect is more efficient in larger grains as RATs are found to be stronger for larger grains \citep{2007MNRAS.378..910L, 2008MNRAS.388..117H}. The observed features in the North and the Center regions may be explained by RAT-A for the increase in $P$ in the regions outside the cores whereas tangled magnetic field along the line-of-sight due to the large column density along the line-of-sight or RAT-D or both may explain the decrease in $P$ for the regions of the embedded cores. We need to disentangle the effect of magnetic field tangling on the depolarization to study the possibility of RAT-D effect in explaining the depolarization at higher dust temperatures. For the South region with no embedded cores and hence the only radiation field is from the diffused ISRF, the increase in $P$ with increasing $T_\mathrm{d}$ may be explained by RAT-A theory.

\subsubsection{Effect of Magnetic Field Tangling} \label{sec:Magnetic Field Tangling}
The observed polarization fraction not only depends on the dust properties and grain alignment but also on the line-of-sight magnetic field geometry. Turbulence of super-Alfvenic in nature in the cloud can cause magnetic field tangling which can result in the decrease of polarization fraction. But to check whether the observed depolarization as the intensity increases is due to the effect of magnetic field tangling or decrease in net alignment efficiency of grains in denser regions or both, we need to disentangle the effect of magnetic field tangling on depolarization. For this, we calculate the polarization angle dispersion function $S$ and $P \times S$. The value of $S$ provides information on the local non-uniformity in the distribution of magnetic field morphology. The value of $P \times S$ provides information on the averaged grain alignment efficiency along the line-of-sight \citep{2020A&A...641A..12P}. For a constant grain alignment efficiency, larger value of $S$ means stronger magnetic field tangling which would result in lower value of polarization fraction and smaller value means weaker magnetic field tangling which would give higher value of polarization fraction.
\begin{figure}[h]
    \centering
    \includegraphics[width=0.47\textwidth]{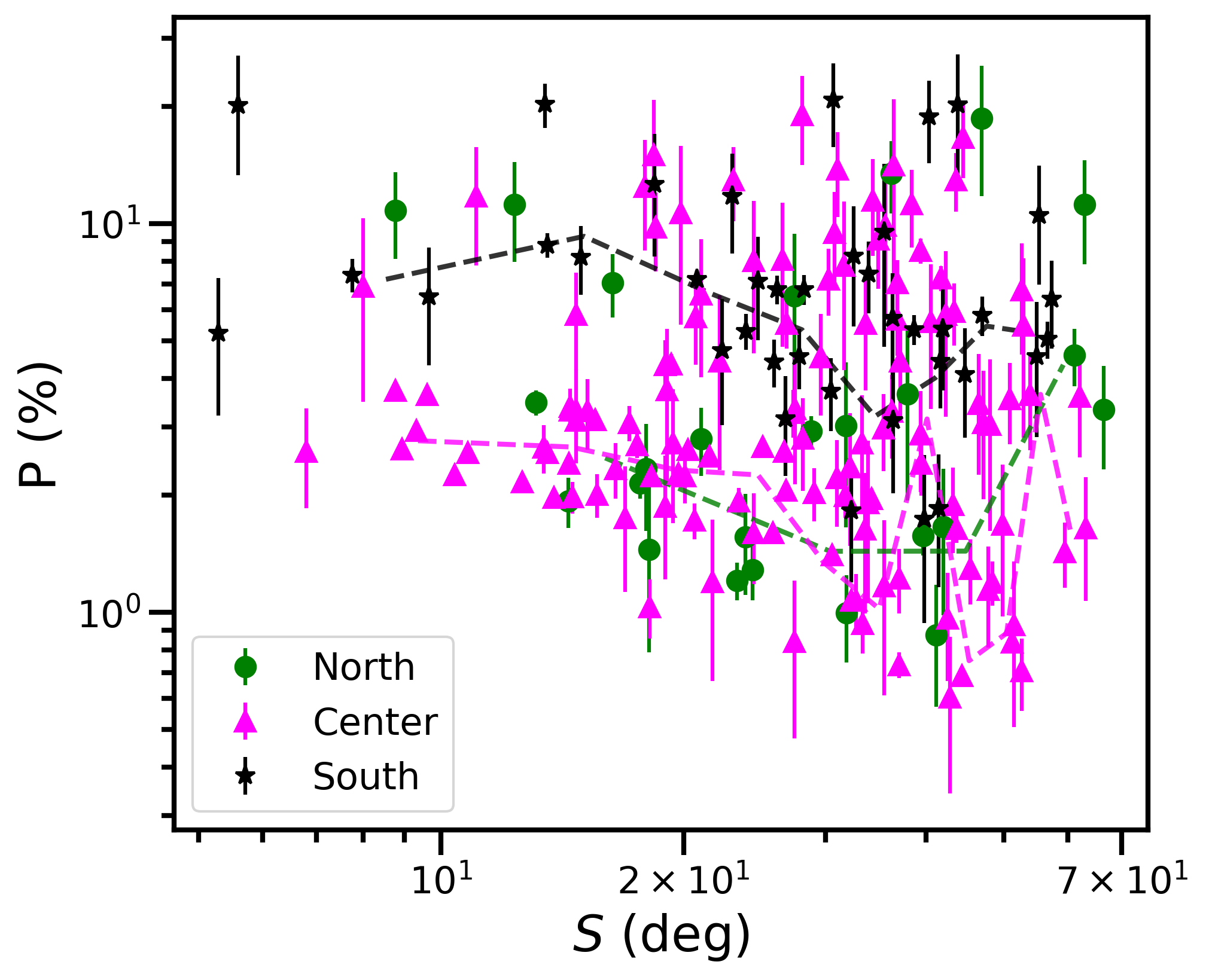}
    \caption{Relation between polarization fraction $P$ and polarization angle dispersion function $S$ in each region.}
    \label{Figure:P_S}
\end{figure}

According to the definition described in Section 3.3 of \cite{2020A&A...641A..12P}, we calculate $S$ using the following relation
\begin{equation}
{
S^2(r,\delta) = \frac{1}{N}\sum\limits_{i=1}^{N} {\left[\psi(r+\delta_i) - \psi(r)\right]}^2,
}
\end{equation}
 where the sum extends over the $N$ pixels, indexed by $i$ and located at positions $r+\delta_i$, within a circle centered on $r$ and having radius of $\delta$ taken as two times the beam size of JCMT/POL-2. The term $[\psi(r+\delta_i) - \psi(r)]$ is the difference in the polarization angles at positions $r+\delta_i$ and $r$.

\begin{figure*}
    \centering
    \includegraphics[width=0.8\textwidth]{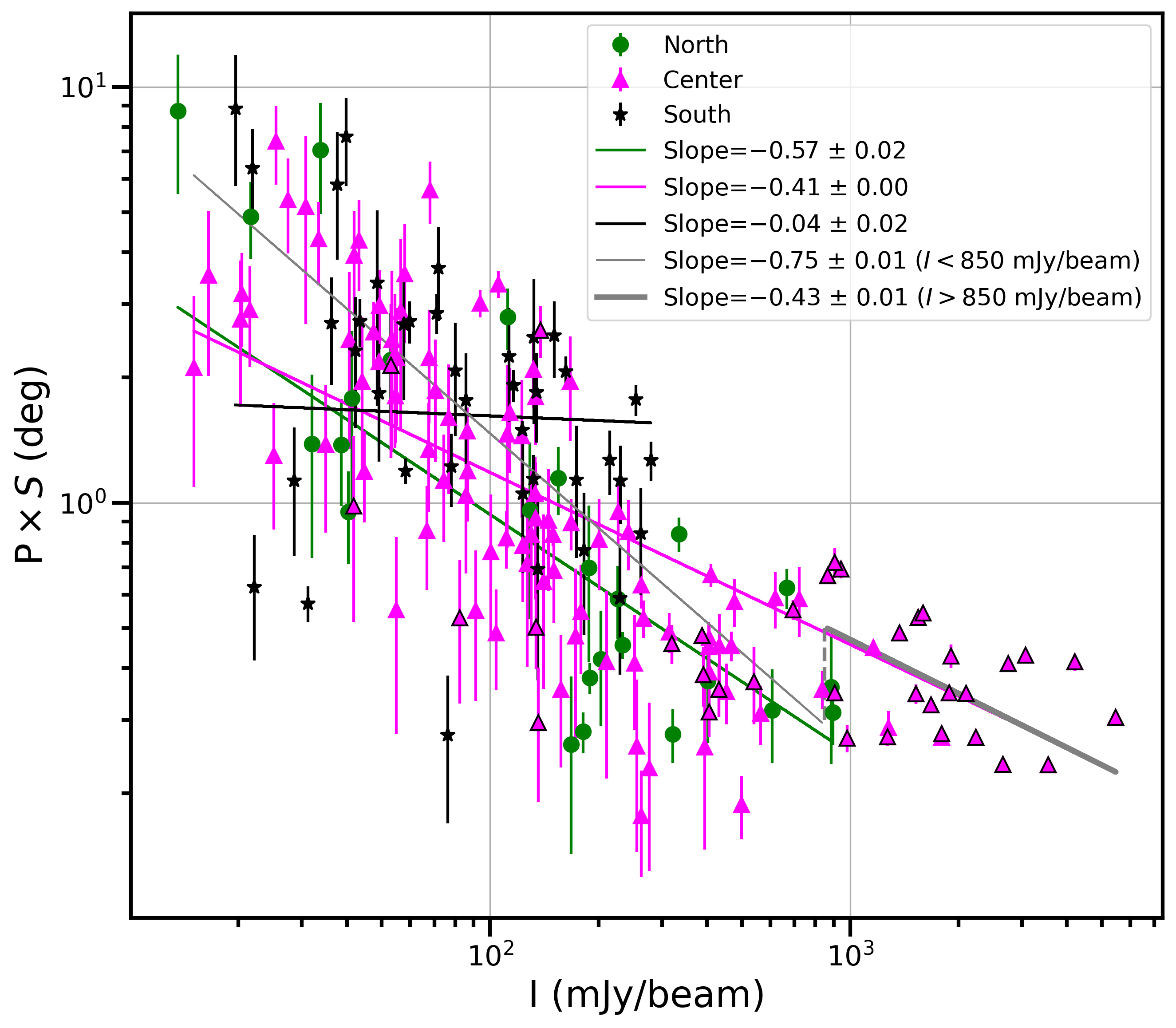}
    \caption{Variation of averaged grain alignment efficiency $P \times S$ with total intensity $I$ in each region. All the notations are the same as in Figure \ref{Figure:P_I}.}
    \label{Figure:P_S_I}
\end{figure*}

As there are noises on Stokes parameters $Q$ and $U$, $S$ is biased. This bias of $S$ can be positive or negative based on whether the true value is smaller or larger than the random polarization angle $52^\circ$ \citep{2016A&A...595A..57A}. An estimate of the variance of $S$ ($\sigma_S$) because of noise and then the debiased $S$ ($S_\mathrm{db}$) as described in Section 3.5 of \cite{2020A&A...641A..12P} are given by

\begin{equation}
\begin{split}
&\sigma_S^2(r,\delta)=\frac{\sigma_{\psi}^2(r)}{N^2S^2}\left[\sum\limits_{i=1}^{N} {\psi(r+\delta_i) - \psi(r)}\right]^2 + \\ &\hspace{1.5cm} \frac{1}{N^2S^2}\sum\limits_{i=1}^{N} \sigma_{\mathrm{\psi}}^2(r+\delta_i)\left[\psi(r+\delta_i) - \psi(r)\right]^2
\end{split}
\end{equation}


and
\begin{equation}
{
S_\mathrm{db}^2(r,\delta) = S^2 - \sigma_S^2 \hspace{0.5cm} \rm{if} \hspace{0.5cm} \it{S > \sigma_S}
}
\end{equation}
We use only those values of $S_\mathrm{db}$ with $S > \sigma_S$ and other values not satisfying this condition are removed. Hereafter, we use $S$ to mean $S_\mathrm{db}(r,\delta)$ for convenience. Then, we study the variation of $P$ with $S$ as shown in Figure \ref{Figure:P_S}. We fit the weighted running means for each region instead of best-fit lines. The data points show large spread and fitting best-fit lines do not reflect well the large spreadness of the data. The weighted running means show more clearly the variations. We find that the North region does not show significant variation of $P$ with $S$. The Center region shows some slight decreasing nature of $P$, however overall there is no much significant correlation between $P$ and $S$. In the South region, we find that $P$ decreases with $S$.

Again, we study the variation of averaged alignment efficiency $P \times S$ with total intensity $I$ as shown in Figure \ref{Figure:P_S_I} and find that the alignment efficiency decreases with intensity in the North and Center regions as similar with variations of $P$ with $I$ in Figure \ref{Figure:P_I} but decreases slightly in the South region. The slope for the South region becomes shallower in $P \times S$ variation with $I$ than $P$ variation with $I$ which may imply that the magnetic field tangling is significant to cause depolarization in the South region and the slight decrease in averaged alignment efficiency with intensity in the South region can be explained by RAT-A. Also, the decrease in averaged alignment efficiency with intensity in the North and Center regions can be explained dominantly by decrease in grain alignment efficiency in denser regions.

\begin{figure*}
    \centering
    \includegraphics[width=0.8\textwidth]{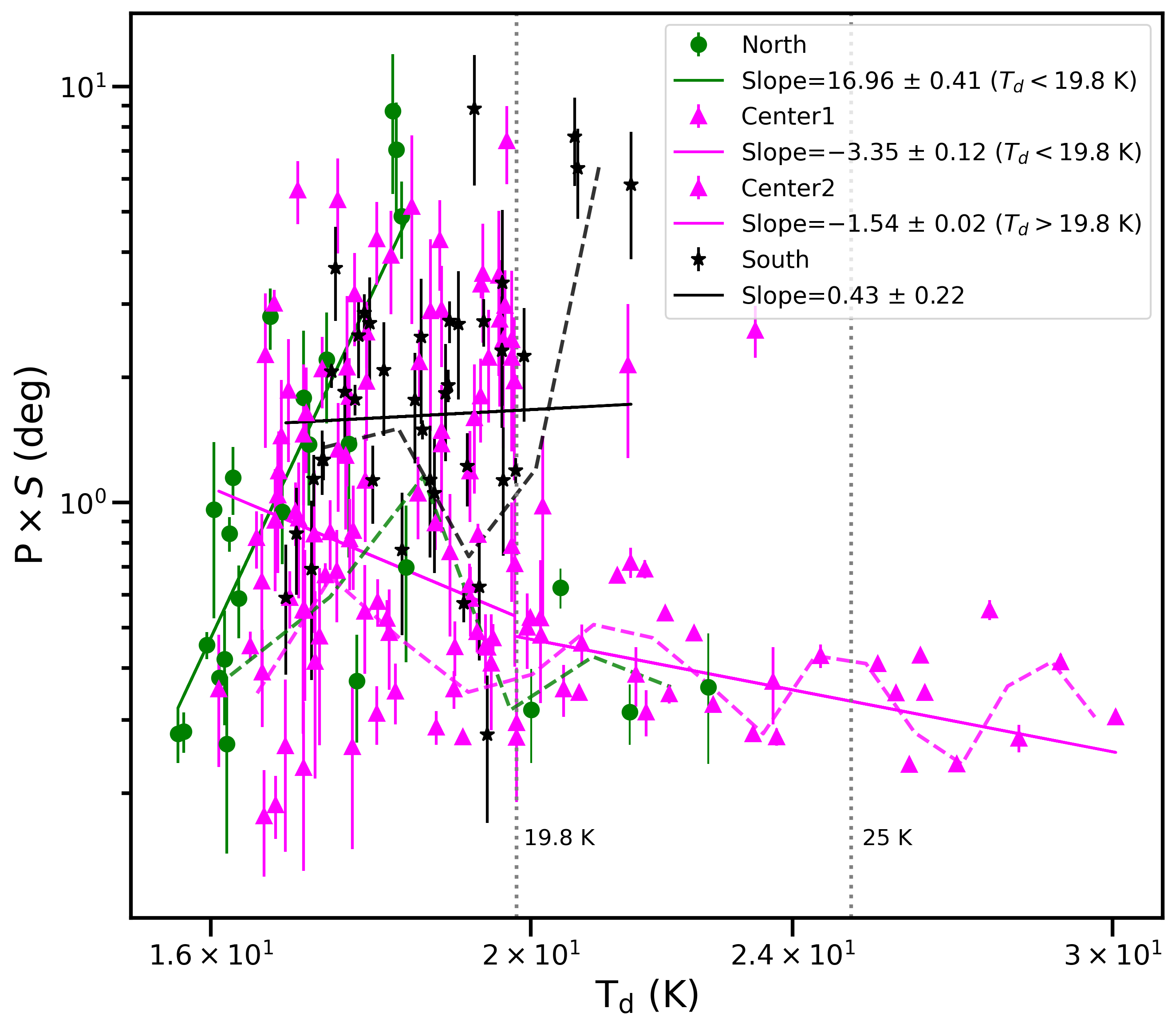}
    \caption{Variations of the averaged grain alignment efficiency $P \times S$ with dust temperature $T_\mathrm{d}$ in each region. All the notations are the same as in Figure \ref{Figure:P_Td}.}
    \label{Figure:P_S_Td}
\end{figure*}

\begin{figure*}
    \centering
    \begin{tabular}{ccc}
        \includegraphics[scale=0.5]{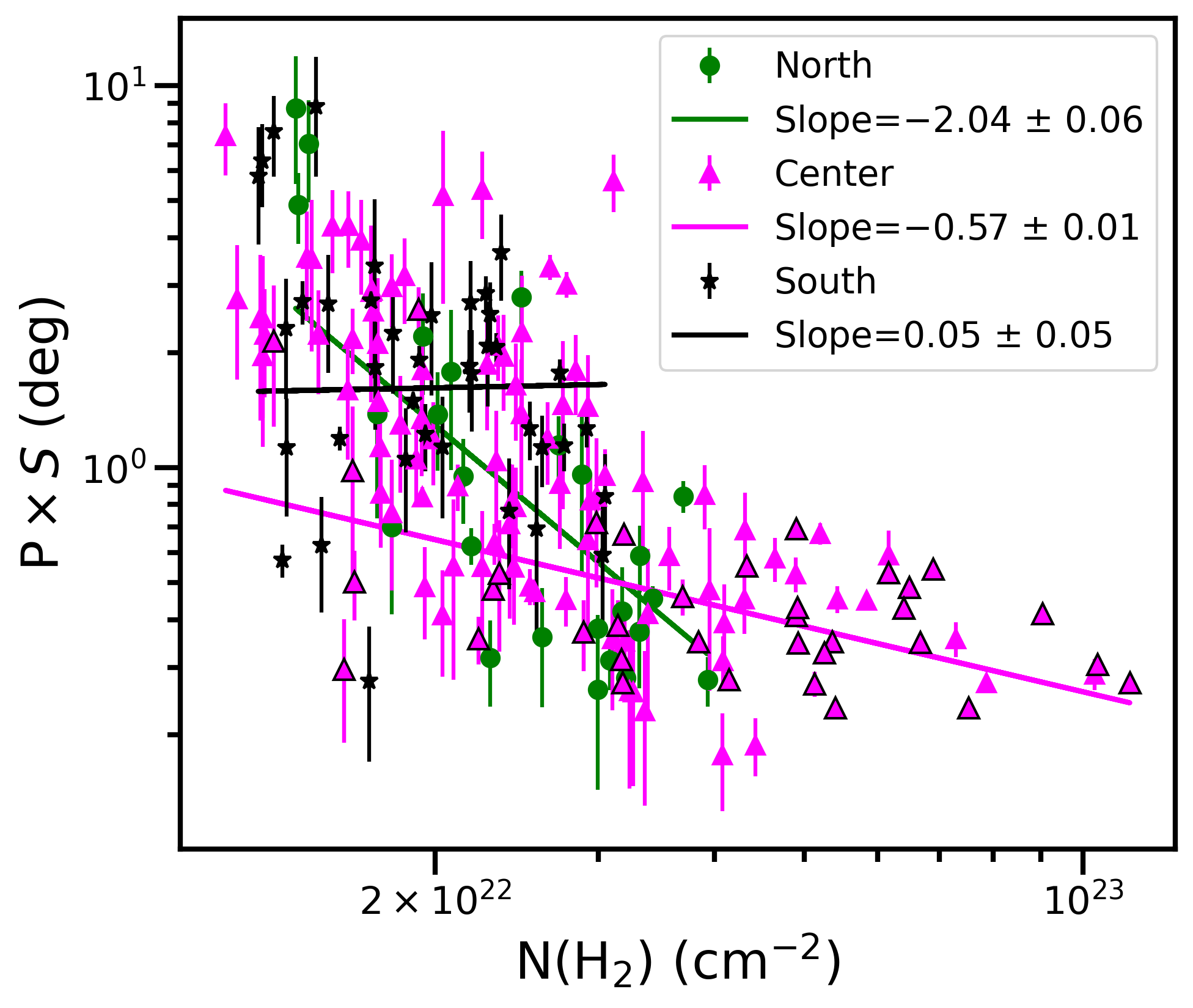} &
        \includegraphics[scale=0.5]{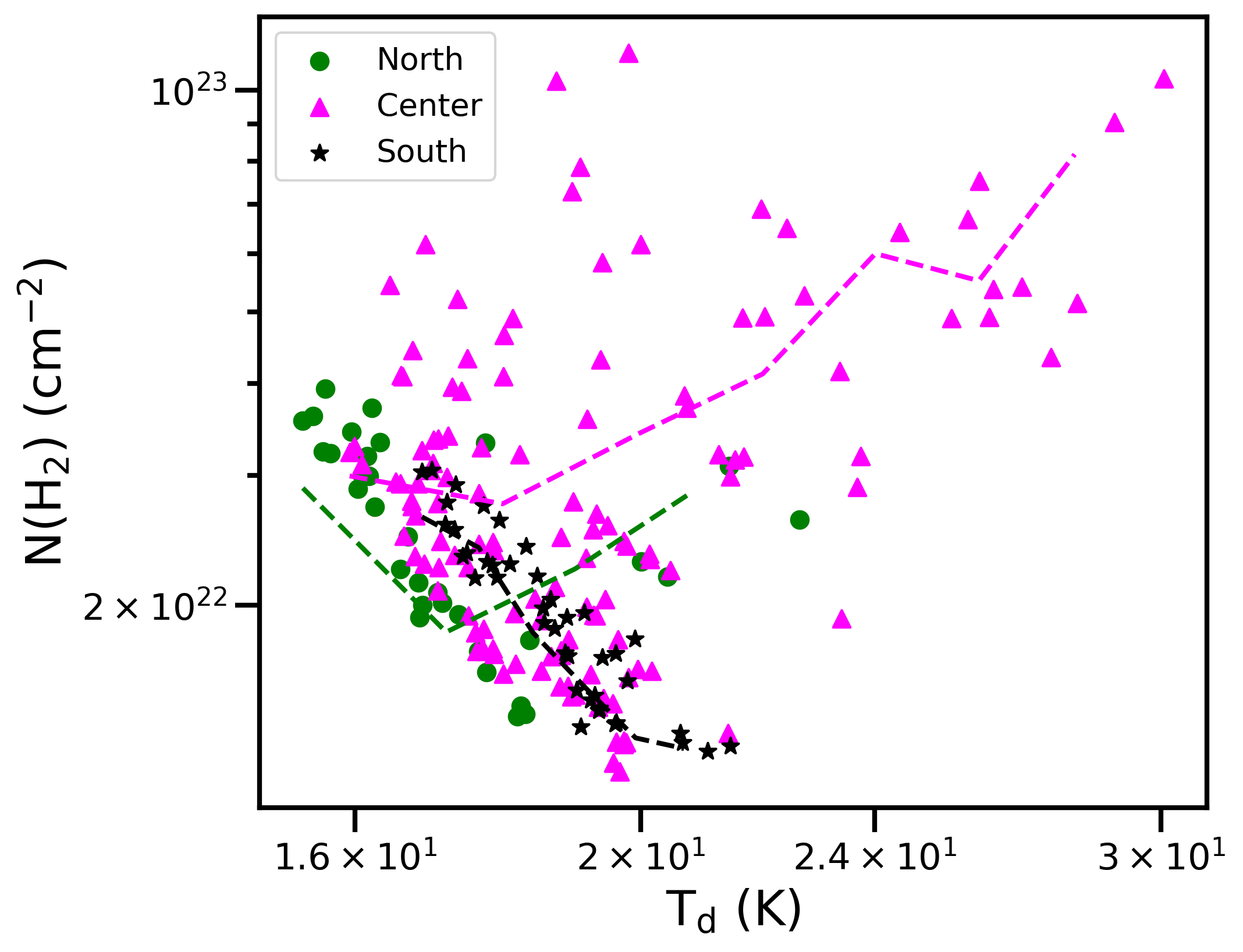}
    \end{tabular}
    \caption{Variations of the averaged grain alignment efficiency $P \times S$ with the gas column density $N(\mathrm{H_2})$ (left) and $N(\mathrm{H_2})$ with dust temperature $T_\mathrm{d}$ (right) in each region.}
    \label{Figure:P_S_NH2}
\end{figure*}

The variation of $P \times S$ with $T_\mathrm{d}$ is shown in Figure \ref{Figure:P_S_Td} and it follows nearly similar to the variation of $P$ with $T_\mathrm{d}$ (see Figure \ref{Figure:P_Td}). The variations of $P \times S$ with $N(\mathrm{H_2})$ and $N(\mathrm{H_2})$ with $T_\mathrm{d}$ are shown in left and right panels of Figure \ref{Figure:P_S_NH2}. In the $N(\mathrm{H_2})-T_\mathrm{d}$ plot, $N(\mathrm{H_2})$ decreases initially and then increases as $T_\mathrm{d}$ increases in the North region which implies that the high dust temperature region of MM3 has high gas column density. Similarly, the Center region containing MM1 and MM2 has high dust temperatures and gas column densities. In the South region which does not have embedded protostellar cores, $N(\mathrm{H_2})$ decreases with $T_\mathrm{d}$. In the variations of $P \times S$ with increasing $T_\mathrm{d}$ and $N(\mathrm{H_2})$, the averaged grain alignment efficiency decreases significantly in the North and the Center regions. In South region, the averaged grain alignment efficiency does not decrease but becomes nearly flat with increasing $N(\mathrm{H_2})$ but $P$ decreases with increasing $N(\mathrm{H_2})$ as shown in Figure \ref{Figure:P_NH2}. Hence, the magnetic field tangling effect is significant in the South region and less significant in the North and the Center regions.    

In the $P \times S$ with $T_\mathrm{d}$ plot (Figure \ref{Figure:P_S_Td}), the averaged alignment efficiency of grains decreases up to 19.8 K in the Center region when the fitting is weighted fit. However, up to 19.8 K there is high spread in the data with a feature of $P$ tending to increase. To study the variations more clearly, we plot weighted running means which show initial increase in $P \times S$ and then decreases at higher $T_\mathrm{d}$ values. After 19.8 K, the spread becomes less and the grain alignment efficiency decreases. In the North region, the grain alignment efficiency increases up to 19.8 K and then decreases. In the South region, the grain alignment efficiency slightly increases in the weighted fit but increases overall in the weighted running means at higher $T_\mathrm{d}$ values. This variation of $P \times S$ with $T_\mathrm{d}$ is nearly similar to the variation of $P$ with $T_\mathrm{d}$ as shown in Figure \ref{Figure:P_Td} for the North and Center regions but the South region shows shallower slope. The magnetic field tangling seems to have some role in causing depolarization in the South region. In the North and Center regions, the decrease in averaged alignment efficiency of grains at larger dust temperature values may imply the possibility of the effect of RAT-D in the protostellar core regions of MM3 (North), MM1 and MM2 (Center) which have high $T_\mathrm{d}$ values due to the presence of high internal radiation fields from the central protostars embedded in these cores.

\subsection{Grain alignment and disruption mechanisms} \label{sec:Grain size}
\subsubsection{Minimum alignment size of grains}
\begin{figure*}
    \centering
    \begin{tabular}{ccc}
        \includegraphics[scale=0.52]{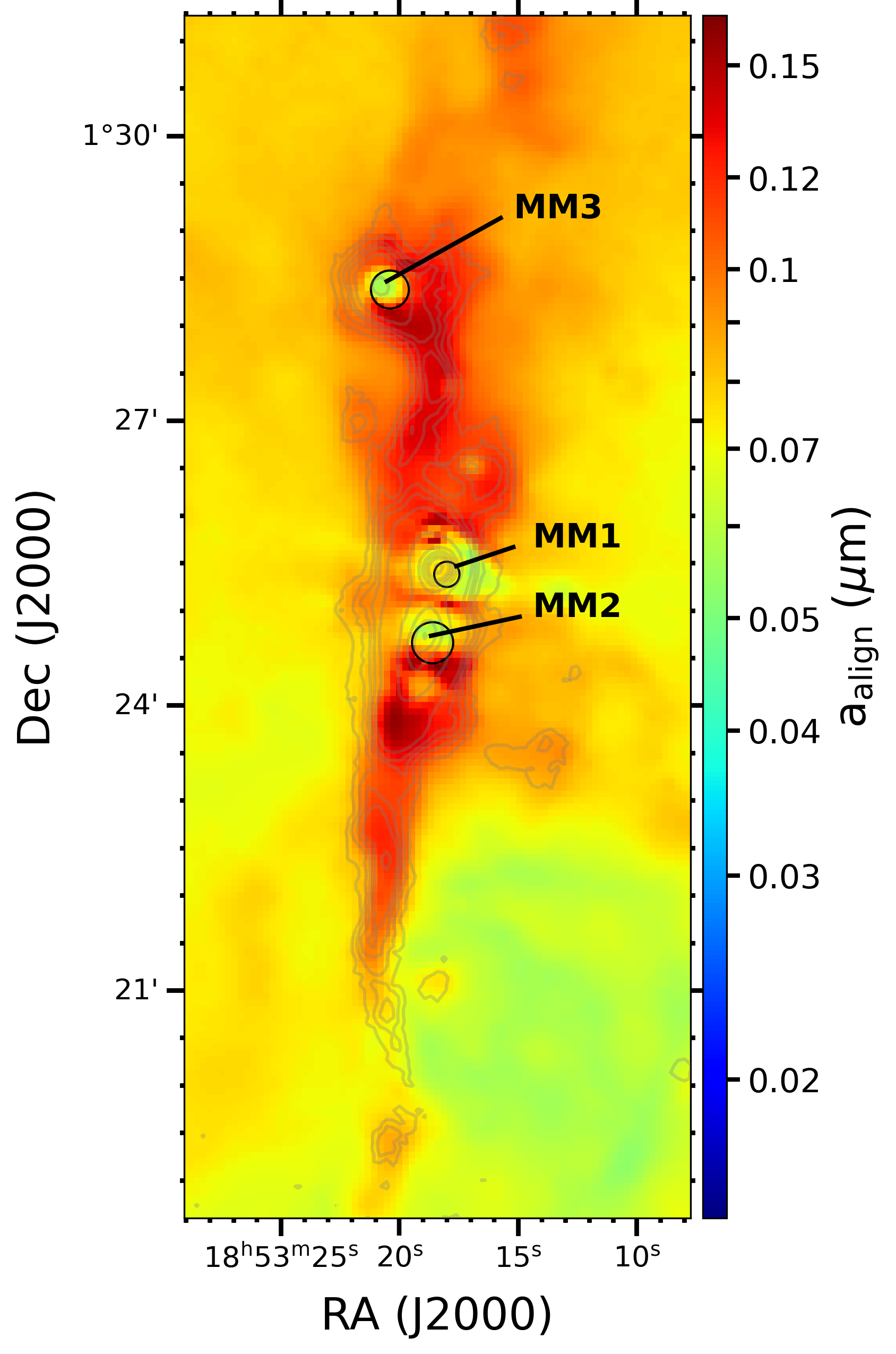} &
        \hspace{10pt} 
        \includegraphics[scale=0.52]{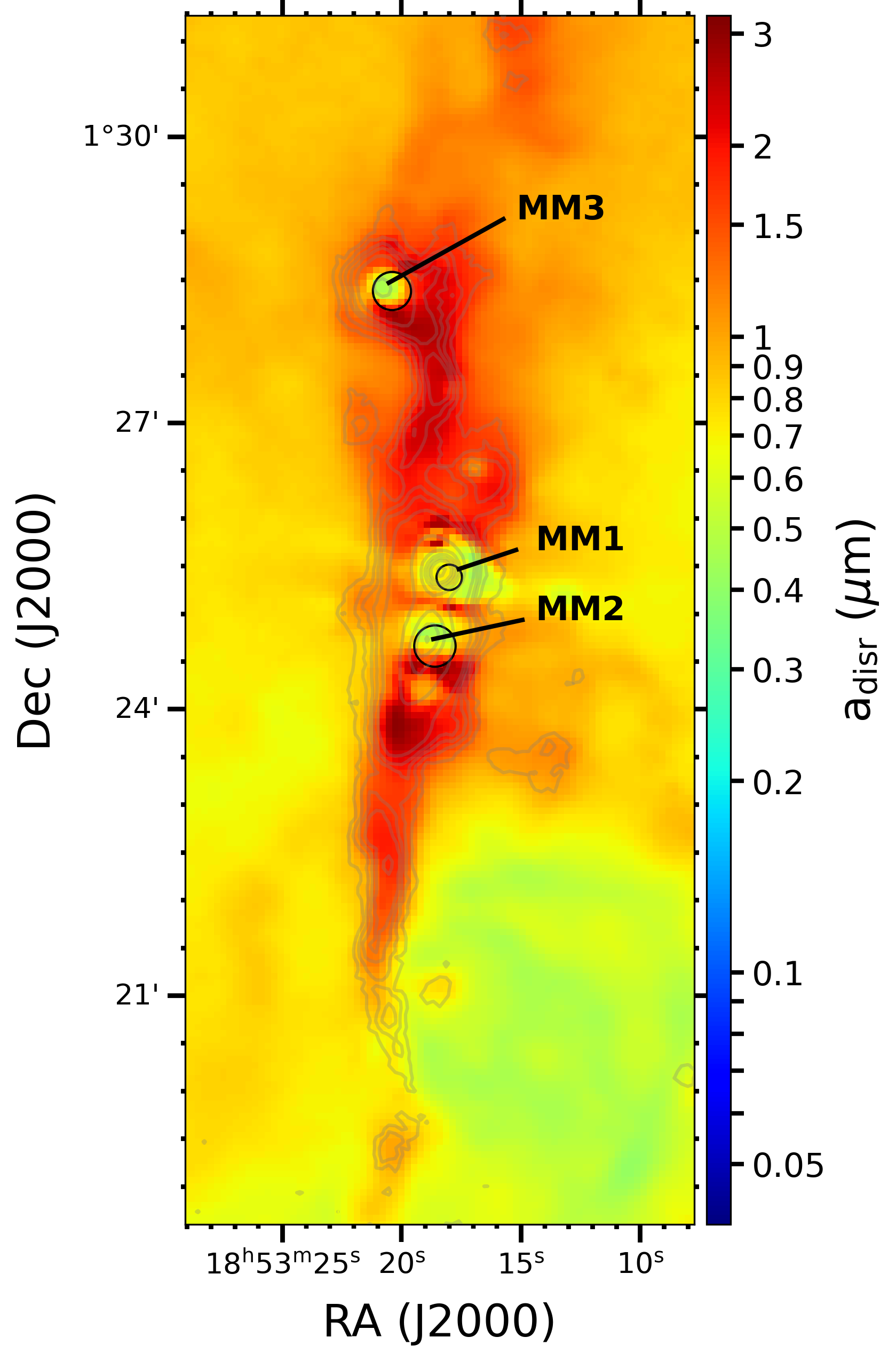} &
    \end{tabular}
    \caption{Maps of minimum alignment size, $a_\mathrm{align}$ (left) and minimum disruption size, $a_\mathrm{disr}$ (right).}
    \label{Figure:a_align_map}
\end{figure*}

The study of grain sizes is very important as far as the RAT theory is concerned in the study of grain alignment mechanisms. From RAT-A theory, effective alignment of grains can be achieved only when they can rotate with a rate far more than the thermal rotation rate i.e suprathermal rotation \citep{2008MNRAS.388..117H, 2016ApJ...831..159H} so that the randomization of grains by gas-grain collisions can be neglected. The dust polarization fraction is determined by the size distribution of aligned grains, spanning from minimum size of aligned grains, hereafter alignment size, $a_{\mathrm{align}}$ to the maximum grain size, $a_{\mathrm{max}}$ \citep{2014MNRAS.438..680H, 2020ApJ...896...44L}. We use the following relation as given in \cite{2021ApJ...908..218H} to calculate the values of $a_{\mathrm{align}}$ in all the regions of the filament.

\begin{equation}
\begin{split}
a_{\mathrm{align}}\simeq0.055\hat{\rho}^{-1/7}\left(\frac{\gamma U}{0.1}\right)^{-2/7}\left(\frac{n_\mathrm{H}}{10^3 \: \mathrm{cm^{-3}}}\right)^{2/7} \\ \times \left(\frac{T_\mathrm{gas}}{10 \: \mathrm{K}}\right)^{2/7} \left(\frac{\bar{\lambda}}{1.2 \: \mu \text{m}}\right)^{4/7} \left(1 + F_\mathrm{IR}\right)^{2/7}, 
\end{split}
\label{equation:a_align} 
\end{equation}
where $\hat{\rho} = \rho_\mathrm{d}/(3$ $\mathrm{gcm^{-3}})$ with $\rho_\mathrm{d}$ being the dust mass density; $\gamma$ is the anisotropy degree of the radiation field; $\bar{\lambda}$ represents the mean wavelength of the radiation; $U$ is the radiation field strength; $n_\mathrm{H}$ is the number density of hydrogen atoms; $T_\mathrm{gas}$ is the gas temperature and $F_\mathrm{IR}$ is the ratio of the IR damping to the collisional damping rate. We take $\gamma = 0.3$ for the regions outside the MM1, MM2 and MM3 cores as the radiation becomes more anisotropic in elongated dense filamentary clouds \citep{1997ApJ...480..633D, 2007ApJ...663.1055B}. However, for the protostellar cores MM1, MM2 and MM3, we take $\gamma = 1$ as the radiation field is dominated by the internal radiations from the high-mass and high-luminosity central protostars and the radiation field becomes unidirectional within these core regions. We use $\rho_\mathrm{d}=3$ $\mathrm{gcm^{-3}}$, $\bar{\lambda}=1.2$ $\mu$m, $n_\mathrm{H} = 2n(\mathrm{H_2})$ where $n(\mathrm{H_2})$ is the volume density of molecular hydrogen gas and $T_\mathrm{gas}=T_\mathrm{d}$ is considered as this thermal equilibrium between gas and dust is valid for dense and cold environments. For dense molecular clouds, $F_\mathrm{IR} << 1$. To calculate $U$, we use the relation between dust temperature and the radiation strength for silicate grains having sizes in the range of 0.01-1$\mu$m with dust heating and cooling balance and radiation strength $U < 10^4$ $(\approx 75$ K) i.e $U$ $\approx$ $(T_\mathrm{d}/16.4$ $\mathrm{K})^6$ \citep{2011piim.book.....D}. The map of the alignment size is given in the left panel of Figure \ref{Figure:a_align_map}. The histogram distribution of the alignment size for each region is shown in Figure \ref{Figure:Histogram_a_align_disr} (left panel). The green and magenta vertical dotted lines denote the median value of $0.08 \pm 0.01$ $\mu$m for the North region with $T_\mathrm{d} > 19.8$ K and of 0.07 $\mu$m for the Center region with $T_\mathrm{d} > 25$ K and the shaded regions denote the uncertainties in these median values, respectively.  We find that the alignment size increases from the outer layer to the inside in the filament for the North except the region of MM3 which shows a significant reduction in alignment size and South regions but for the region containing MM1 and MM2 in the Center region, the alignment size significantly decreases.   

In the context of the RAT paradigm, aligned grains have their size distributions ranging from $a_\mathrm{align}$ to $a_\mathrm{max}$ and the polarization fraction is determined by this range of grain size distributions. The value of $a_\mathrm{max}$ is determined by grain growth and destruction processes. When a certain $a_{\mathrm{max}}$ is given, increasing in the value of $a_{\mathrm{align}}$ can result in narrower size distribution of aligned grains which can reduce polarization fraction $P$. Also, decreasing in the value of $a_{\mathrm{align}}$ can result in wider size distribution of aligned grains which can give higher $P$ (see Figure 7 in \citealt{2022FrASS...9.3927T}). So, it is expected to get an anti-correlation between $a_{\mathrm{align}}$ and $P$. Also, it is expected that $a_\mathrm{align}$ should increase with $I$ in starless clouds.

\begin{figure}[h]
    \centering
        \includegraphics[scale=0.5]{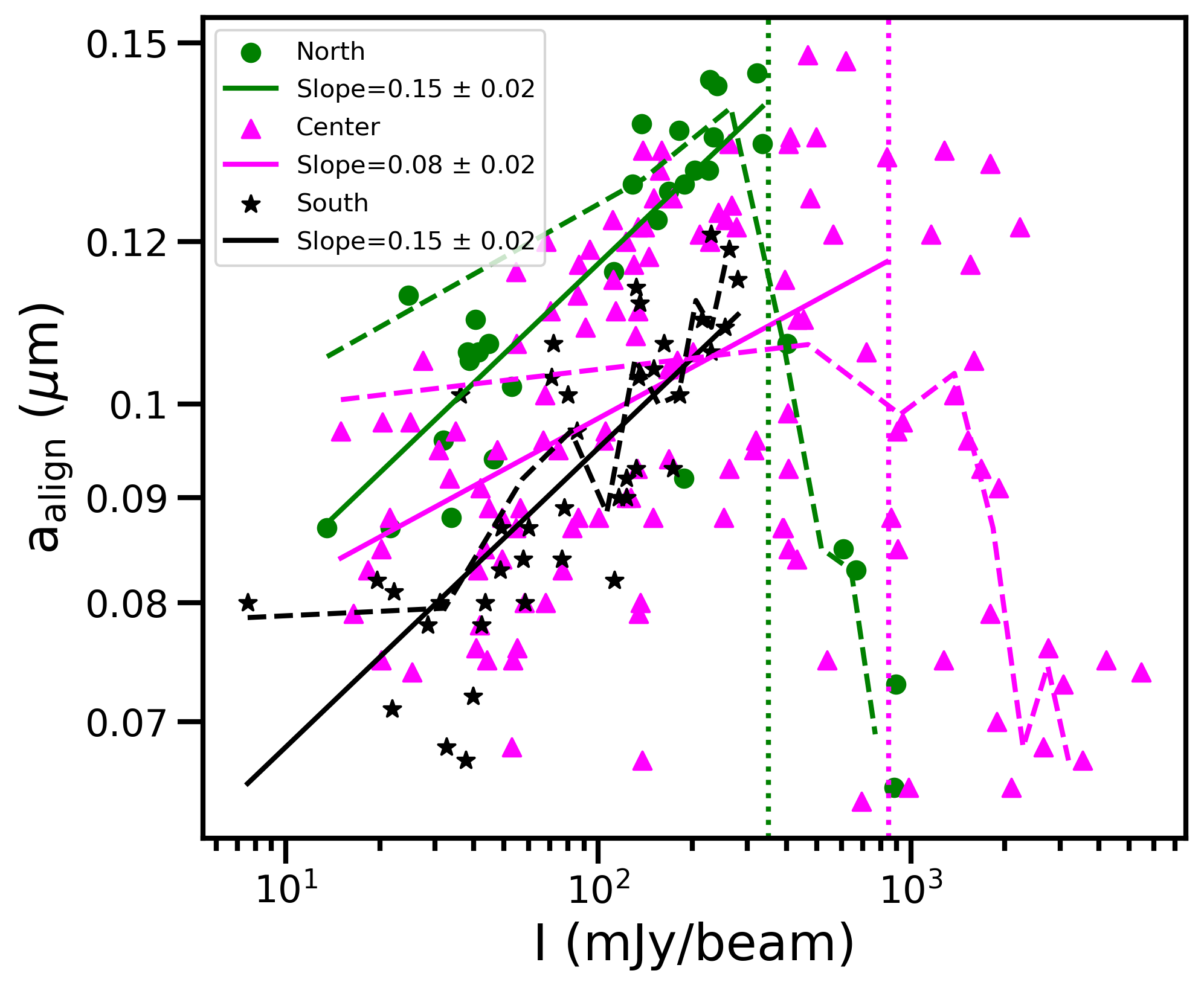} 
    \caption{Variation of $a_{\mathrm{align}}$ with $I$. The dashed lines are the running means and the solid lines are the best-fit lines of the respective regions. The best-fit lines for the North and the Center regions are fitted only up to $I=350$ mJy/beam and $I=$ 850 mJy/beam marked with vertical green and magenta dotted lines, respectively.}
    \label{Figure:a_I}
\end{figure}

\begin{figure*}
    \centering
    \begin{tabular}{ccc}
        \includegraphics[scale=0.5]{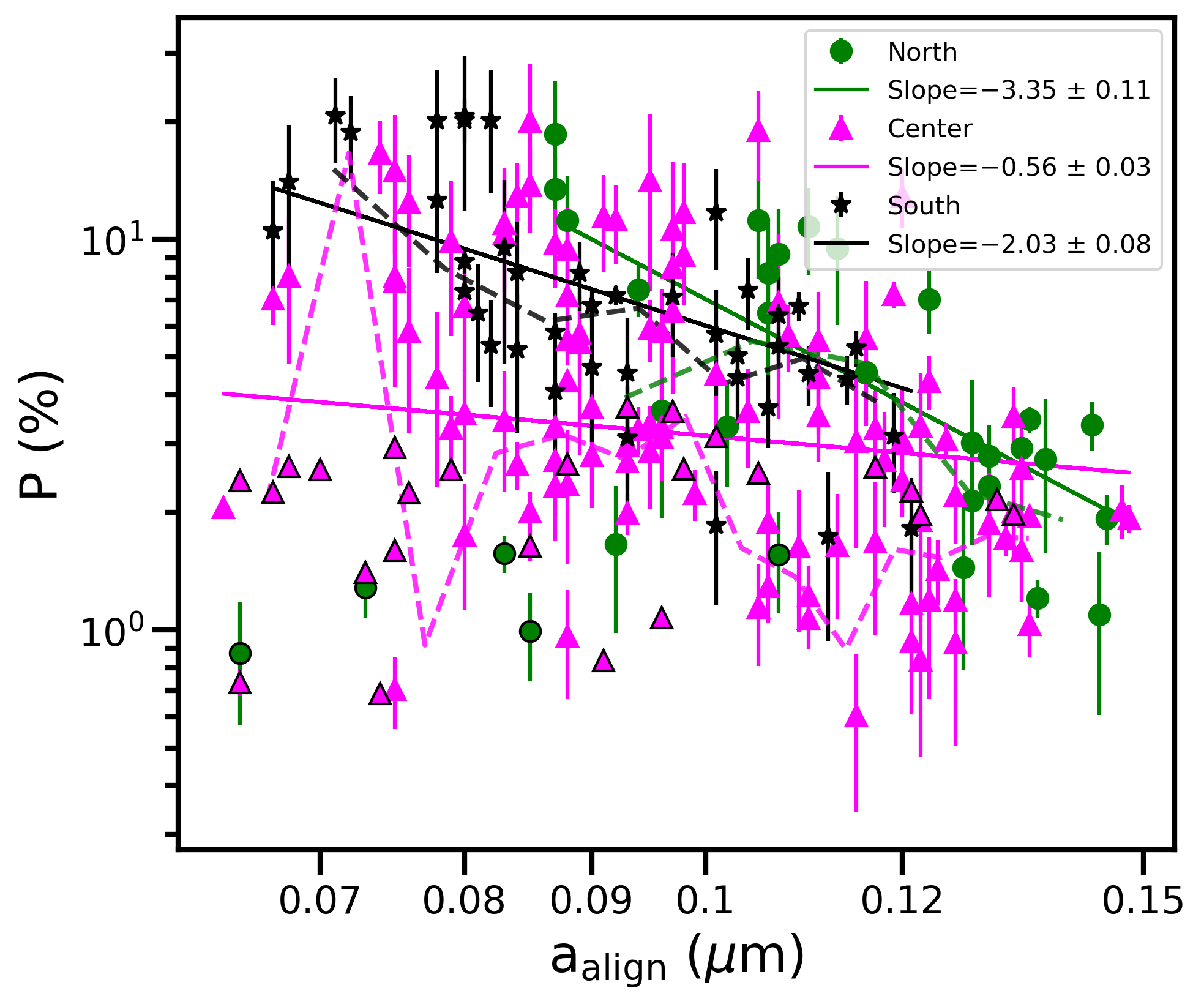} & 
        \hspace{5pt}
        \includegraphics[scale=0.5]{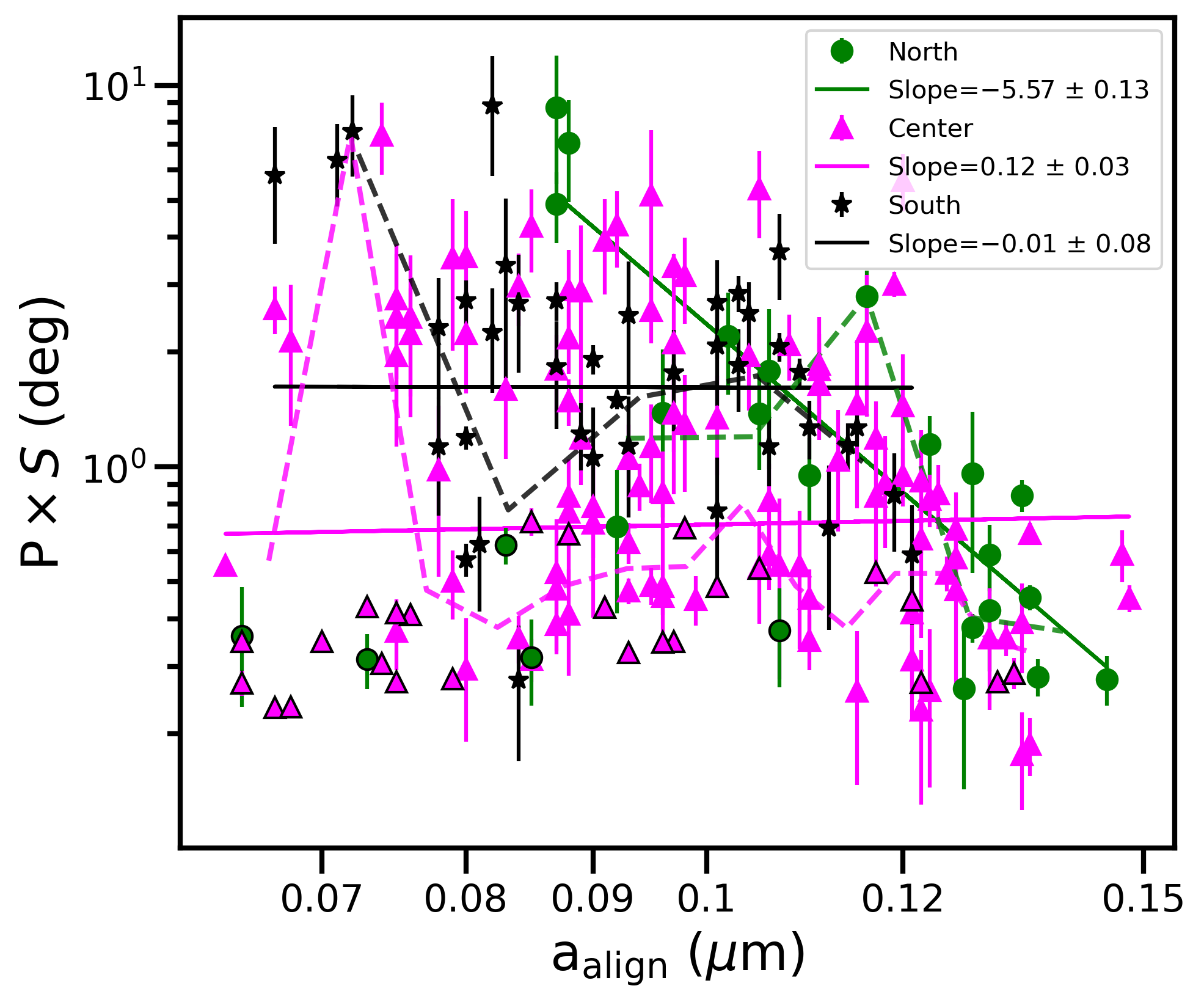} &
    \end{tabular}
    \caption{Variations of $P$ with $a_{\mathrm{align}}$ (left) and $P \times S$ with $a_{\mathrm{align}}$ (right) in each of the regions. The dashed lines are the weighted running means and the solid lines are the weighted best-fit lines. For the North and the Center regions, the running means and the fits are done only for $I < 350$ mJy/beam and $I < 850$ mJy/beam, respectively. The data points after these $I$ values are shown with black edgecolors.}
    \label{Figure:P_PS_a}
\end{figure*}

Figure \ref{Figure:a_I} shows the variation of $a_{\mathrm{align}}$ with total emission intensity $I$. The North and South Regions show nearly similarly an increase in $a_{\mathrm{align}}$ as $I$ increases up to 350 mJy/beam which means that $a_{\mathrm{align}}$ increases in denser regions and then the North region shows decrease in $a_\mathrm{align}$ which can be due to the presence of dense and hot core MM3.
The Center region shows an increase in $a_{\mathrm{align}}$ at first up to around 850 mJy/beam but at higher intensity values $a_{\mathrm{align}}$ decreases which can be due to the presence of internal radiations from the cores MM1 and MM2. The modeling of grain alignment and grain disruptions in very dense protostellar core as given in \cite{2021ApJ...908..218H} shows that because of the effect of the stellar radiation field $a_{\mathrm{align}}$ first changes slowly in the envelope and then decreases rapidly when entering the region around the protostar due to the increase in radiation flux and constant gas density. In the presence of these protostars the grains can be disrupted by RATs into smaller fragments and it can decrease $P$. Figure \ref{Figure:P_PS_a} shows the variation of $P$ (left) and $P \times S$ (right) with $a_{\mathrm{align}}$. We plot weighted power-law fit and also weighted running means for the regions up to 350 mJy/beam for the North region and up to 850 mJy/beam for the Center region. The data points after these values are indicated with black edgecolors.

It is found that the value of polarization fraction $P$ and averaged grain alignment efficiency $P \times S$ decreases in the North region significantly and shows some spread in the Center and South regions but decreases overall at higher $a_{\mathrm{align}}$ values. Polarization fraction decreases as $a_{\mathrm{align}}$ increases due to the reduction in the fraction of aligned grains, from numerical modeling \citep{2020ApJ...896...44L, 2021ApJ...908..218H}. These features in the North and the Center regions may be associated with the regions outside the cores. The other data points shown in blackedgecolors that correspond to $I > 350$ mJy/beam for the North region and $I > 850$ mJy/beam for the Center region show small $P$ and $P \times S$ values overall. At smaller $a_{\mathrm{align}}$ values, we expect higher polarization fraction but the $P$ and $P \times S$ values are small even at smaller $a_{\mathrm{align}}$ values for these intensity values. These data points may be associated with the regions of the cores. The decrease in the polarization fraction at smaller $a_{\mathrm{align}}$ values is opposite to the expectation of RAT-A theory. Therefore, we expect the possibility of RAT-D effect in these protostellar cores.

\subsubsection{Minimum disruption size of grains}
The depolarization with increasing $T_\mathrm{d}$ may also be caused by the fluctuations in the average B-field orientations along the line-of-sight within the unresolved cores. There have been many studies to investigate the B-fields in G34 filament at different wavelengths with dust and line emission polarization observations. For example, the mapping of the B-field morphologies in the regions of MM1 and MM2 cores have been done using dust polarization observations by JCMT/POL-2 at 850 $\mu$m with $14''.1$ resolution \citep{2019ApJ...883...95S}, TADPOL/CARMA\footnote{Combined Array for Research in Millimeter-wave Astronomy} at 1.3 mm with $2''.5$ resolution \citep{2014ApJS..213...13H} and SMA\footnote{Sub-Millimeter Array} at 870 $\mu$m with $1''.5$ resolution \citep{2014ApJ...792..116Z}. A comparison of the B-field morphologies mapped by these different observations in the MM1 and MM2 core regions is given in Figure 9 of \cite{2019ApJ...883...95S}. The SMA observations show deviation in the B-field orientations from the JCMT/POL-2 observations in these cores. However, in all these observations, the B-field orientations in these core regions are almost uniform. Therefore, the contribution from the magnetic field tangling along the line-of-sight on the observed depolarization in these dense cores is expected to be small. The depolarization with increasing $T_\mathrm{d}$ in the North and Center regions may also be caused by B-field tangling and/or change in the B-field inclination angle within the unresolved beam \citep{2024ApJ...965..183H}. However, there lacks the details on the B-field tangling and inclination variation. Here, we explore whether RAT-D can happen in these warm environments associated with the cores MM1, MM2 in the Center and MM3 in the North regions and explain the observed depolarization by estimating the minimum disruption size of grains. Since the MM1, MM2 cores have high bolometric luminosities of the order of $10^4$ $L_\odot$ with masses 433 $M_\odot$ and 533 $M_\odot$, respectively and MM3 of the order of nearly $10^3$ $L_\odot$ with mass 171 $M_\odot$ which may imply the presence of high-mass protostars \citep{2010ApJ...715..310R}, the possibility of occurring RAT-D in these hot and dense cores is expected to be high.

As mentioned in the introduction in Section \ref{section:introduction}, large grains when exposed to very high radiation field could be disrupted into smaller fragments when the centrifugal stress induced by rapid rotation exceeds the tensile strength of the grains, termed as  RAT-D mechanism. The grain disruption depends on their tensile strength, denoted by $S_{\mathrm{max}}$ which is an uncertain parameter determined by their structure and composition. Composite grains have $S_\mathrm{max} \approx 10^6-10^7$ erg $\mathrm{cm^{-3}}$ and compact grains have $S_\mathrm{max} \approx 10^9-10^{11}$ erg $\mathrm{cm^{-3}}$ \citep{2019NatAs...3..766H}. In dense molecular clouds, grains are expected to be large and have composite structures due to grain coagulation processes. The disruption of grains by RAT-D mechanism decreases the observed polarization fraction due to depletion of large grains. We calculate the minimum size of grains that can be disrupted known as disruption size, $a_\mathrm{disr}$ using the analytical formula given in \cite{2021ApJ...908..218H} as follows

\begin{equation}
\begin{split}
a_{\mathrm{disr}}\simeq1.7\left(\frac{\gamma_{-1}U}{n_3 T_1^{1/2}}\right)^{-1/2} \left(\frac{\bar{\lambda}}{1.2 \: \mu \text{m}}\right) \hat{\rho}^{-1/4} S_{\mathrm{max,7}}^{1/4} \\ \times \left(1 + F_\mathrm{IR}\right)^{1/2} \mu \text{m}
\end{split}
\label{equation:a_disr}
\end{equation}

\begin{figure*}
    \centering
    \begin{tabular}{ccc}
        \includegraphics[scale=0.52]{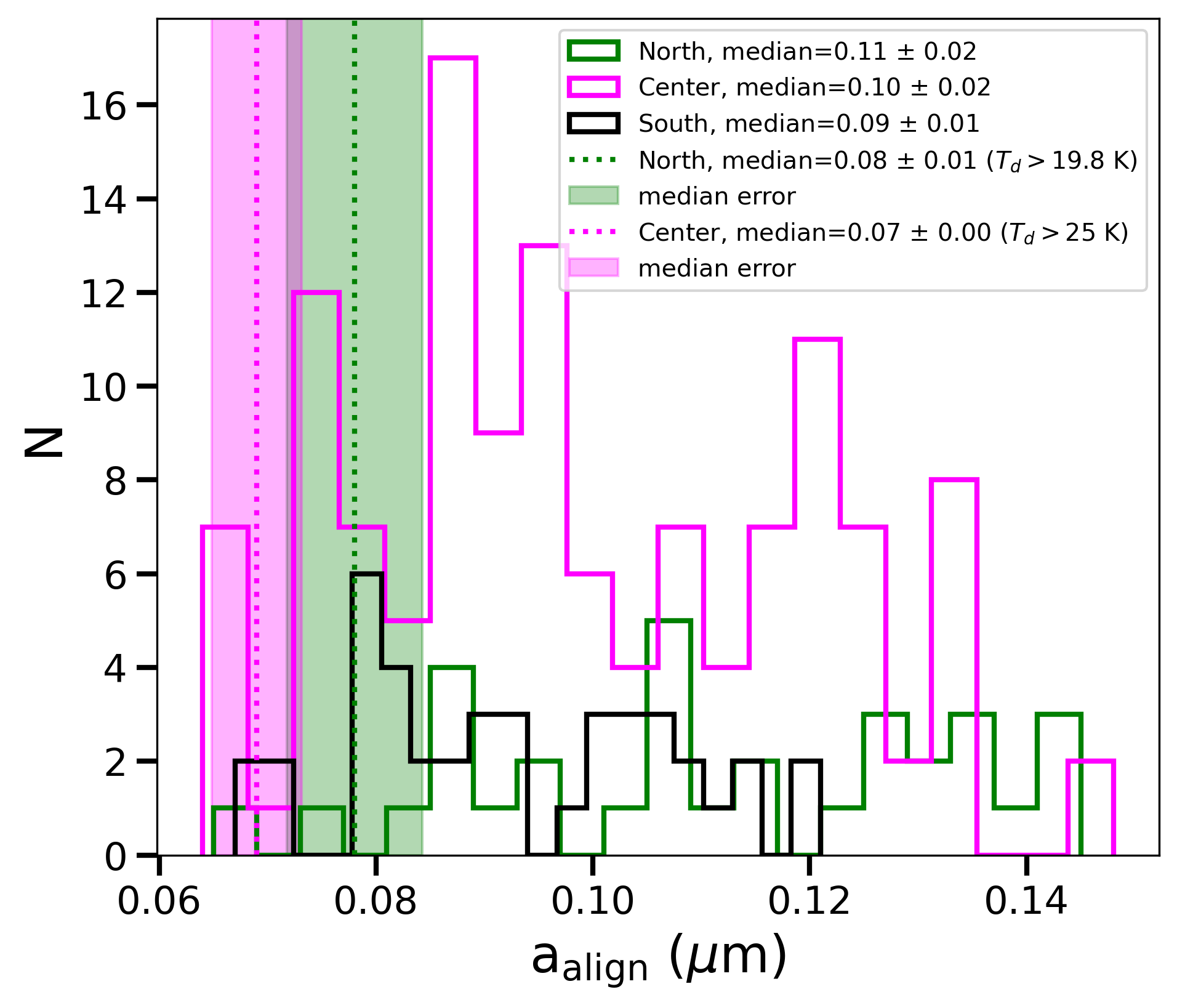} & 
        \hspace{5pt}
        \includegraphics[scale=0.52]{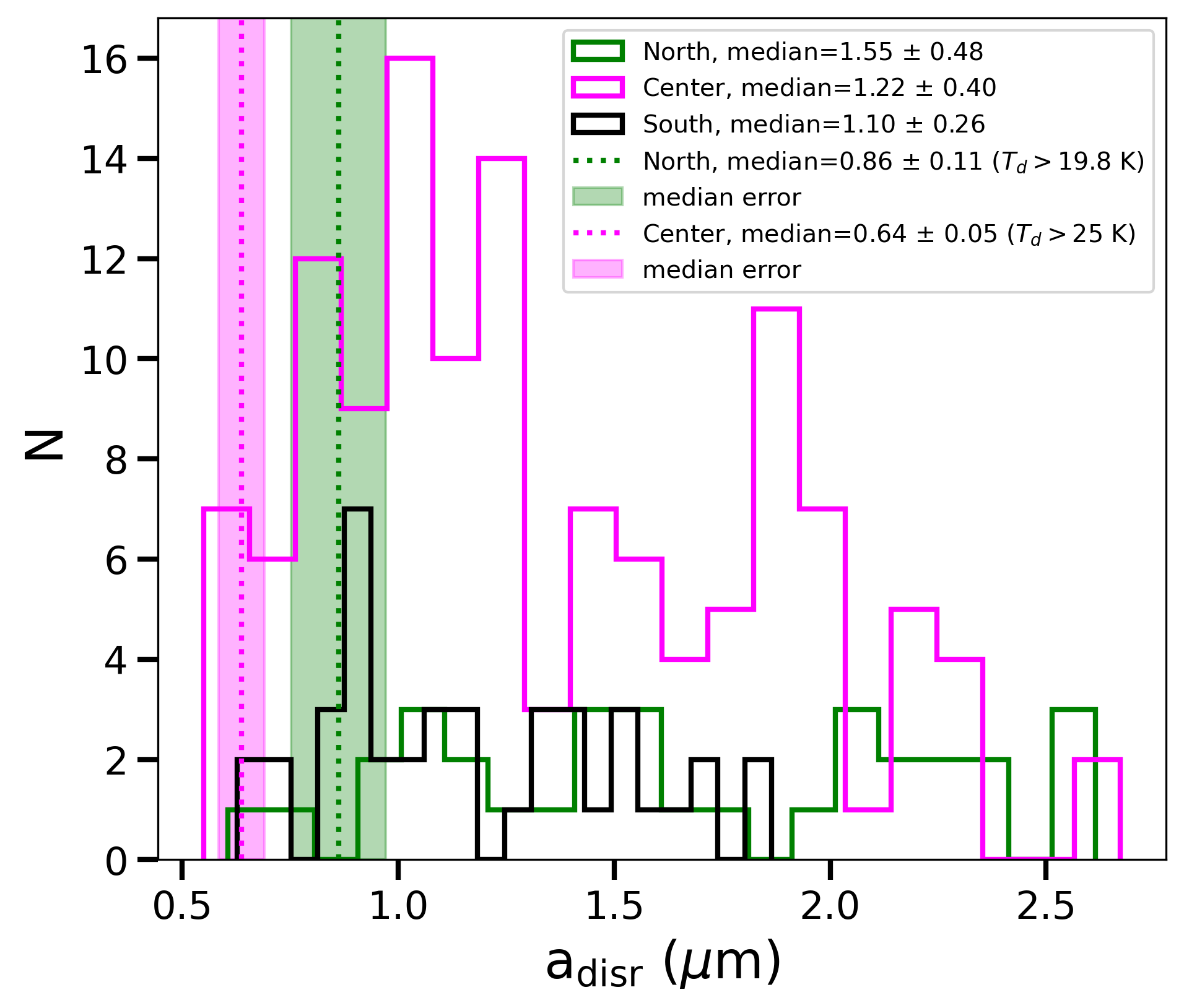} &
    \end{tabular}
    \caption{Histogram distributions of $a_{\mathrm{align}}$ (left panel) and $a_{\mathrm{disr}}$ (right panel) for each of the North, Center and South regions. The vertical green and magenta dotted lines in each panel represent the median values for the North region with $T_\mathrm{d} > 19.8$ K and the Center region with $T_\mathrm{d} > 25$ K and the shaded regions the corresponding uncertainties in the median values, respectively.}
    \label{Figure:Histogram_a_align_disr}
\end{figure*}

The $a_\mathrm{disr}$ value depends on the local gas properties, the strength of radiation field and the grain tensile strength. For our calculation of $a_\mathrm{disr}$, we consider $S_\mathrm{max} = 10^5$ erg $\mathrm{cm^{-3}}$ for large porous composite grains that are expected in dense regions of the filament due to grain coagulation processes. From the numerical simulations for porous grain aggregates by \cite{2019ApJ...874..159T}, the parameter $S_\mathrm{max}$ can be calculated using the analytical formula given in Equation 33 in \cite{2019ApJ...874..159T}. For large composite grains made up of monomers (constituent particles) of radius 0.1 $\mu$m, volume filling factor of 0.1 and surface energy per unit area of the material of $10^2$ erg $\mathrm{cm^{-2}}$, $S_\mathrm{max} \approx 10^5$ erg $\mathrm{cm^{-3}}$. The other parameter values are taken the same as those used in the calculation of $a_\mathrm{align}$ in Equation \ref{equation:a_align}. The map of the minimum disruption size thus calculated is shown in the right panel of Figure \ref{Figure:a_align_map}. The histogram distribution of the minimum disruption size of grains for each region is shown in Figure \ref{Figure:Histogram_a_align_disr} (right panel). The green and magenta vertical dotted lines denote the median value of $0.86 \pm 0.11$ $\mu$m for the North region with $T_\mathrm{d} > 19.8$ K and of $0.64 \pm 0.05$ $\mu$m for the Center region with $T_\mathrm{d} > 25$ K and the shaded regions show the uncertainties in these median values, respectively. The regions of MM3 (North), MM1 and MM2 (Center) show a significant decrease in $a_\mathrm{disr}$ values which reach up to nearly 0.6 $\mu$m at high dust temperatures and the other regions of the filament show the $a_\mathrm{disr}$ values above 1 $\mu$m.

\begin{figure*}
    \centering
    \includegraphics[width=1\textwidth]{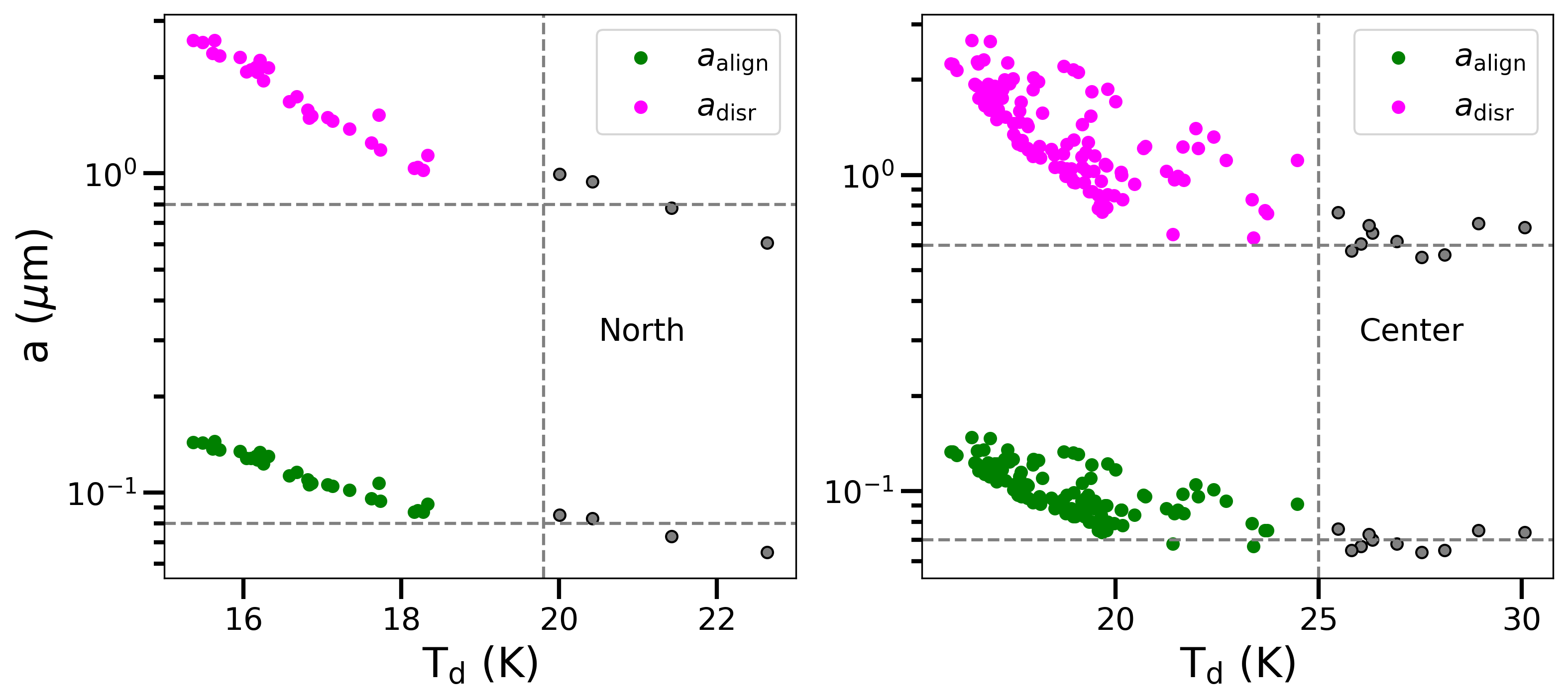}
    \caption{Grain size distributions for RAT-A and RAT-D mechanisms as a function of dust temperature in North (left) and Center (right) regions. The gray points correspond to $T_\mathrm{d} > 19.8$ K for the North and $T_\mathrm{d} > 25$ K for the Center regions. The vertical gray dotted lines are drawn at $T_\mathrm{d} = 19.8$ K and 25 K which act as the transition temperatures (see Figure \ref{Figure:P_Td})  and the horizontal lines are drawn at a = 0.08 $\mu$m and a = 0.8 $\mu$m for North that indicate the median $a_\mathrm{align}$ and $a_\mathrm{disr}$ values for $T_\mathrm{d} > 19.8$ K, respectively and at a = 0.07 $\mu$m and a = 0.6 $\mu$m for the Center region for $T_\mathrm{d} > 25$ K.}
    \label{Fig:a_Td}
\end{figure*}

In Figure \ref{Fig:a_Td} which shows the variations of $a_\mathrm{align}$ (green) and $a_\mathrm{disr}$ (magenta) as a function of dust temperatures for the North (left panel) and the Center (right panel) regions with the gray points denoting the respective values after 19.8 K for the North and after 25 K for the Center regions which act as the transition temperatures (for Center region, though the P decreases after 19.8 K the more significant decrease in P occurs after around 25 K) (see Figure \ref{Figure:P_Td}), we find that the disruption and the alignment sizes are large for $T_\mathrm{d} < 19.8$ K in the North and $T_\mathrm{d} < 25$ K in the Center regions but decrease and become nearly constant after 19.8 K and 25 K, respectively. When $T_\mathrm{d} > 19.8$ K, the median value of $a_\mathrm{align} \approx 0.08 \pm 0.01$ $\mu$m and median value of $a_\mathrm{disr} \approx 0.86 \pm 0.11$ $\mu$m in the North region. When $T_\mathrm{d} > 25$ K in the Center region, the median value of $a_\mathrm{align} \approx 0.07$ $\mu$m and of $a_\mathrm{disr} \approx 0.64 \pm 0.05$ $\mu$m. Therefore, larger grains above 0.86 $\mu$m in the North for the regions of $T_\mathrm{d} > 19.8$ K and above 0.64 $\mu$m for the regions of $T_\mathrm{d} > 25$ K in the Center regions can be disrupted thereby resulting in narrower size distributions of grains that can be aligned and hence reduction in the polarization fraction. Also, $P$ and $P \times S$ decrease with $T_\mathrm{d}$ for $T_\mathrm{d} > 19.8$ K and $T_\mathrm{d} > 25$ K in the North and Center regions, respectively (see Figures \ref{Figure:P_Td}, \ref{Figure:P_S_Td}). This can imply the possibility of RAT-D in the dense and hot core regions of the North and Center regions (see Section \ref{section:RAT-D Mechanism} for the discussion).

\subsection{Effect of Magnetic Relaxation on the RAT alignment} \label{sec:Magnetic Relaxation}
In the study of grain alignment, the magnetic properties of dust are very important as they enable the grains to interact with the external magnetic field. When iron atoms are distributed diffusely within a silicate grain, the grain acts as an ordinary paramagnetic material, but when distributed as clusters, the grain becomes super-paramagnetic \citep{2022AJ....164..248H}. A paramagnetic grain rotating with an angular velocity $\omega$ in an external magnetic field $B$ experiences paramagnetic relaxation \citep{1951ApJ...114..206D} that induces the dissipation of the grain rotational energy into heat, resulting in the gradual alignment of angular velocity and angular momentum with $B$. This classical Davis-Greenstein mechanism can be applied to any magnetic material. However, the paramagnetic relaxation alone cannot produce efficient alignment of grains due to gas randomization. Also, only RATs cannot produce perfect alignment of grains because of the dependence of RAT alignment efficiency on various parameters like the angle between the direction of radiation and the magnetic field, the shape of the grains and their compositions \citep{2016ApJ...831..159H, 2021ApJ...913...63H}. Therefore, to explain the observed high polarization fraction of 8-20\% in the filament, it is important to consider super-paramagnetic grains having embedded iron atoms as clusters. However, this high $P$ value in the outer low-surface brightness regions may also be due to missing flux in these extended regions due to observing mode. For a study on this missing flux, please see Appendix \ref{section:synthetic intensity map}. The observed high $P$ is also possible by the modern grain alignment theory that includes alignment by magnetic relaxation. Hence, we explore the effect of magnetic relaxation strength on the RAT alignment efficiency of grains. 

To describe the aligning effect of magnetic relaxation relative to the disalignment by gas collisions, a dimensionless parameter $\delta_\mathrm{mag}$ which gives the strength of the magnetic relaxation was introduced \citep{2016ApJ...831..159H} and is defined as the ratio of the gas collision damping timescale, $\tau_\mathrm{gas}$ and the magnetic relaxation time, $\tau_\mathrm{mag,sp}$. For super-paramagnetic grains having embedded iron clusters which are expected in dense regions because of grain evolution, the strength of magnetic relaxation is given by
\vspace{0.2cm}
\begin{equation}
{
\delta_\mathrm{mag,sp} = \frac{\tau_\mathrm{gas}}{\tau_\mathrm{mag,sp}} = 56a^{-1}_{-5} \frac{N_\mathrm{cl} \phi_\mathrm{sp,-2} \hat{p}^2 B_3^2}{\hat{\rho} n_4 T_\mathrm{gas,1}^{1/2}} \frac{k_\mathrm{sp}(\Omega)}{T_\mathrm{d,1}}, \label{equation:magnetic relaxation}
}
\end{equation} 
where $a_{-5}=a/(10^{-5}$ cm), $B_3=B_\mathrm{tot}/(10^3$ $\mu$G), $n_4=n_\mathrm{H}/(10^4$ $\mathrm{cm^{-3}}$) with $n_\mathrm{H} \approx 2n(\mathrm{H_2})$ for molecular gas, $T_\mathrm{gas,1}=T_\mathrm{gas}/(10$ K), $T_\mathrm{d,1}=T_\mathrm{d}/(10$ K), $\hat{p}=p/5.5$ with $p \approx 5.5$ the coefficient describing the magnetic moment of an iron atom, $N_\mathrm{cl}$ gives the number of iron atoms per cluster, $\phi_\mathrm{sp}$ is the volume filling factor of iron clusters with $\phi_\mathrm{sp,-2}=\phi_\mathrm{sp}/0.01$ and $k_\mathrm{sp}(\Omega)$ is the function of the grain rotation frequency $\Omega$ which is of order unity \citep{2022AJ....164..248H}.

\begin{figure*}
    \centering
    \begin{tabular}{ccc}
        \includegraphics[scale=0.43]{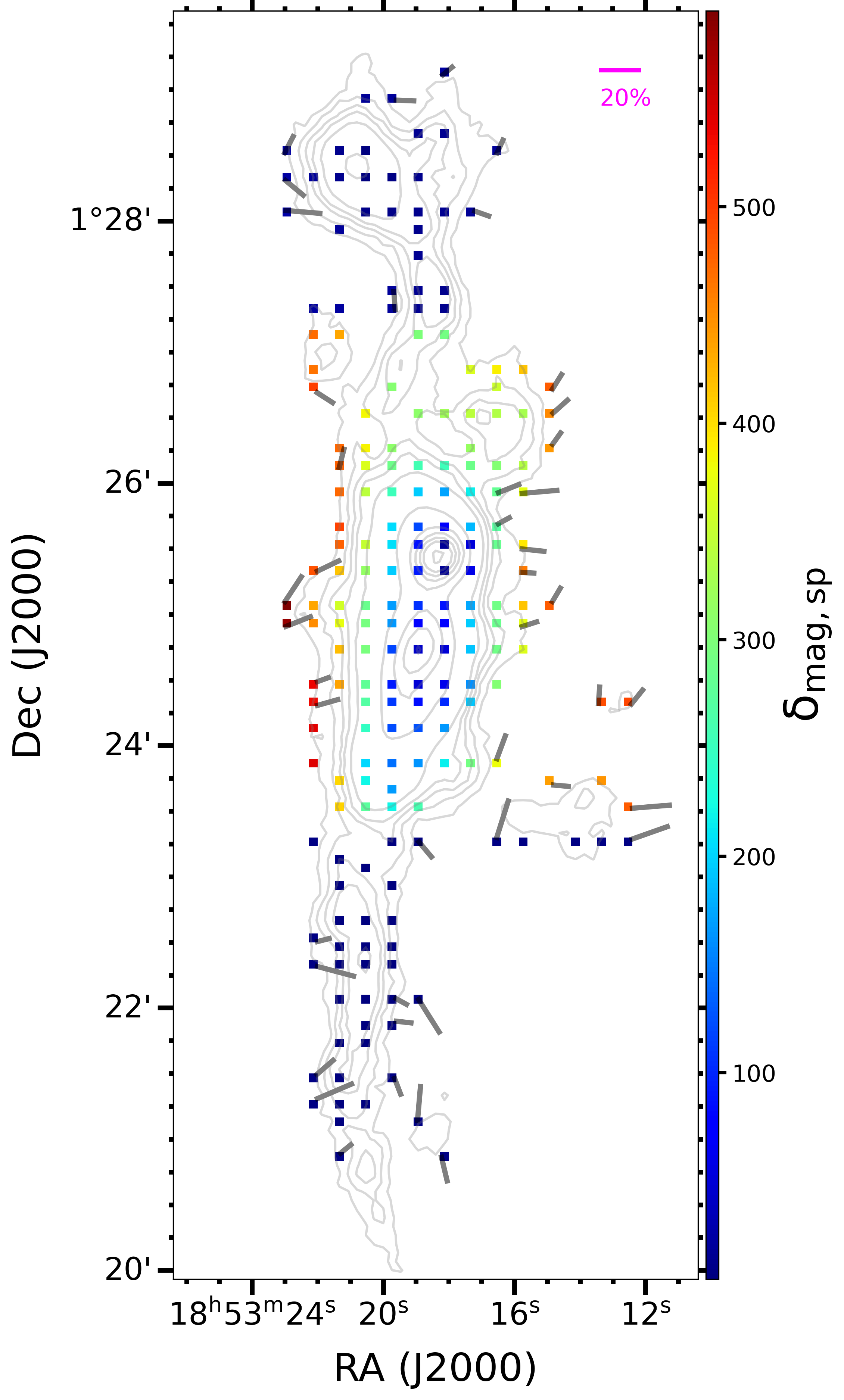} & 
        \hspace{5pt}
        \includegraphics[scale=0.43]{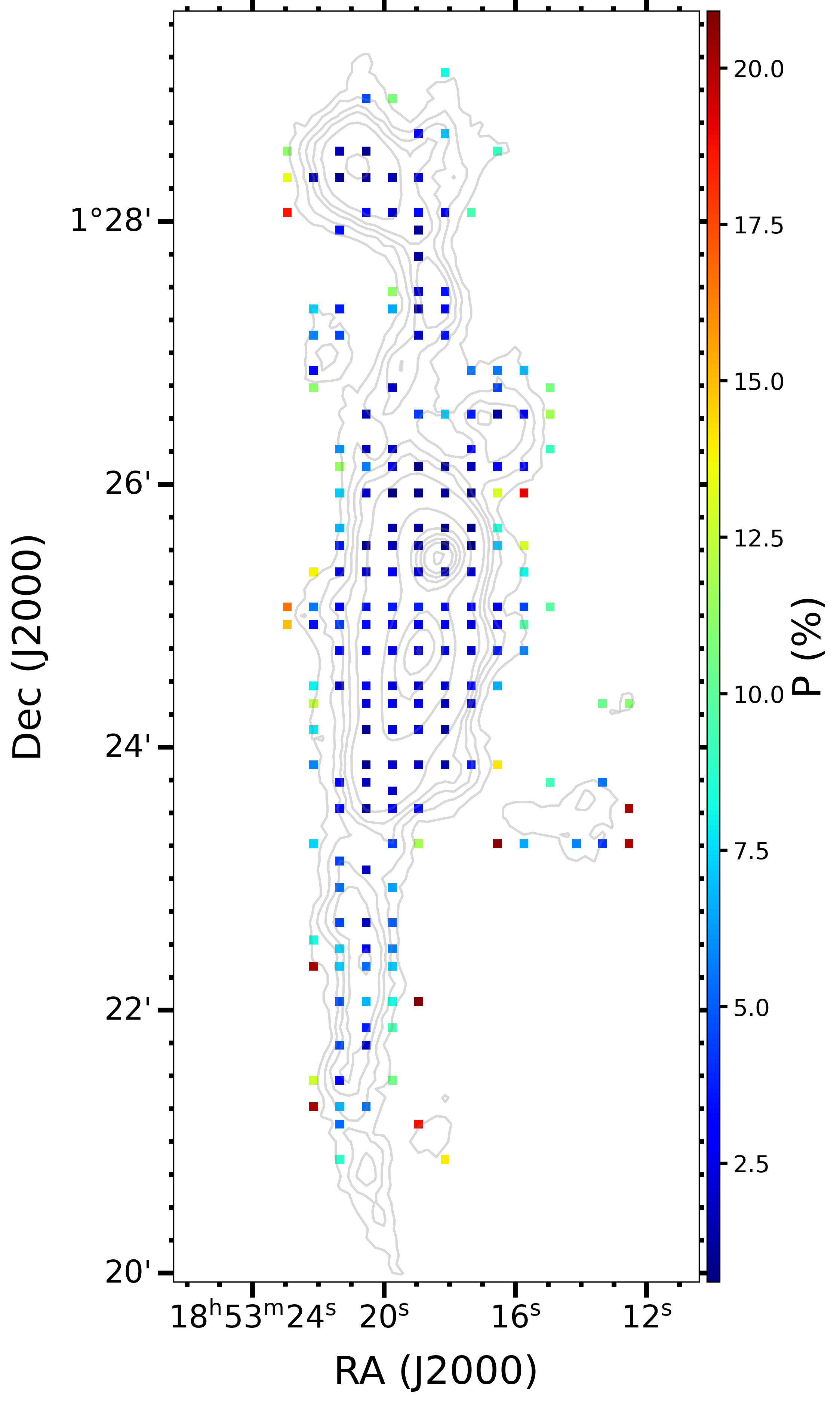} &
    \end{tabular}
    \caption{Maps of the strength of magnetic relaxation, $\delta_\mathrm{mag,sp}$ (left panel) and polarization fraction $P$ (right panel) in all regions of the G34 filament. The contours are the same as in Figure \ref{Figure:Intensity_map}. The gray vectors denote $P$ values ranging from 8-20\%. A reference scale length of $P$ value is also shown. In the outer parts of the filament, especially in the outer parts of the Center region, the values of $\delta_\mathrm{mag,sp}$ are very high with some pixels showing $P$ value from 8-20\%.}
    \label{Fig:mr_map}
\end{figure*}

The magnetic relaxation is considered effective for grain alignment when it happens much faster than the gas collision damping. The combined effect of suprathermal rotation of grains by RATs and magnetic relaxation can enhance the degree of RAT alignment, which is known as the magnetically enhanced radiative torque (MRAT) mechanism. Super-paramagnetic grains can achieve perfect alignment by the MRAT mechanism for $\delta_\mathrm{mag,sp} > 10$ as derived from numerical calculations in \cite{2016ApJ...831..159H}. 

To study the effect of magnetic relaxation on the RAT alignment efficiency in this filament, we calculate the $\delta_\mathrm{mag,sp}$ values in all regions of the filament using Equation \ref{equation:magnetic relaxation}. We take the mean values of plane-of-sky magnetic field strength $B_\mathrm{POS}$ for the North, Center and South regions from \cite{2019ApJ...883...95S} and the values are $100 \pm 40$ $\mu$G, $470 \pm 190$ $\mu$G and $60 \pm 34$ $\mu$G, respectively. The total magnetic field strength $B_\mathrm{tot}$ is 1.3 times the plane-of-sky magnetic field strength $B_\mathrm{POS}$ as given in \cite{2004ApJ...600..279C}. We calculate the $B_\mathrm{tot}$ values for each of the regions and using the values of other parameters we calculate the $\delta_\mathrm{mag,sp}$ values. We use $N_\mathrm{cl}=100$ and $\phi_\mathrm{sp}=0.01$ (about 3\% of iron abundance as iron clusters, \citealt{2016ApJ...831..159H}). The maps of the strength of magnetic relaxation thus calculated and the polarization fraction are shown in Figure \ref{Fig:mr_map}. In the North, the values of $\delta_\mathrm{mag,sp}$ are in the range from $9-25$ with a median value of 14.66, in the Center region the values are in the range from $35-590$ with a median value of 293.09 and in the South from $5-9$ with a median value of 6.32. It is found that the Center region has the strongest magnetic relaxation strength with $\delta_\mathrm{mag,sp} >> 10$. The North region also has $\delta_\mathrm{mag,sp} > 10$. But the South region has $\delta_\mathrm{mag,sp} < 10$. The magnetic field strengths are very large in the Center followed by the North and small in the South regions (see Table \ref{table:magnetic relaxation}). In the outer envelopes of the Center region, there are some pixels with very high magnetic relaxation strength and $P$ values ranging from 8-20\%. Also, in the North region, there are some pixels with $\delta_\mathrm{mag,sp} > 10$ and $P$ values more than 8\%.

\setlength{\tabcolsep}{0.6cm} 
\renewcommand{\arraystretch}{1.6}

\begin{table*}[ht]
    \centering
        \caption{Slope values of weighted power-law fits for different analyses using $PI/\sigma_{PI} > 2$}
    \begin{tabular}{|c|c|c|c|}
        \hline
        
        Relation between & North & Center & South\\ \hline
        $P$ vs. $I$ & $-0.58 \pm 0.02$ & \parbox{3.8cm}{$-0.22 \pm 0.00$ (overall $I$) \\ $-0.54 \pm 0.01$ ($I < 850$ mJy/beam) \\ $-0.36 \pm 0.01$ ($I > 850$ mJy/beam)} & $-0.34 \pm 0.02$ \\ \hline
        $P$ vs. $N(\mathrm{H_2})$ & $-1.18 \pm 0.05$ & \parbox{3.8cm}{$-0.40 \pm 0.01$ (overall $N(\mathrm{H_2})$) \\ $-0.83 \pm 0.01$ ($N(\mathrm{H_2}) < 5 \times 10^{22}$ $\mathrm{cm^{-2}}$) \\ $-1.04 \pm 0.01$ ($N(\mathrm{H_2}) > 5 \times 10^{22}$ $\mathrm{cm^{-2}}$)} & $-1.13 \pm 0.04$ \\ \hline
        $P \times S$ vs. $I$ & $-0.57 \pm 0.02$  & \parbox{3.8cm}{$-0.41 \pm 0.00$ (overall $I$) \\ $-0.75 \pm 0.01$ ($I < 850$ mJy/beam) \\ $-0.43 \pm 0.01$ ($I > 850$ mJy/beam)} & $-0.04 \pm 0.02$  \\ \hline
        $P$ vs. $T_\mathrm{d}$ & 9.82 $\pm$ 0.34 ($T_\mathrm{d} < 19.8$ K) & \parbox{3.7cm}{0.20 $\pm$ 0.12 ($T_\mathrm{d} < 19.8$ K) \\ $-1.58 \pm 0.02$ ($T_\mathrm{d} > 19.8$ K)}  &  5.42 $\pm$ 0.21 \\ \hline
        $P \times S$ vs. $T_\mathrm{d}$  & 16.96 $\pm$ 0.41 ($T_\mathrm{d} < 19.8$ K) & \parbox{3.7cm}{$-3.35 \pm 0.12$ ($T_\mathrm{d} < 19.8$ K) \\ $-1.54 \pm 0.02$ ($T_\mathrm{d} > 19.8$ K)} & 0.43 $\pm$ 0.22  \\ \hline
        $a_\mathrm{align}$ vs. $I$ & \parbox{3.7cm}{0.15 $\pm$ 0.02 \\ ($I < 350$ mJy/beam)} & \parbox{3.7cm}{0.08 $\pm$ 0.02 \\ ($I < 850$ mJy/beam)} & 0.15 $\pm$ 0.02   \\ \hline
        $P$ vs. $a_\mathrm{align}$ & \parbox{3.7cm}{$-3.35 \pm 0.01$ \\ ($I < 350$ mJy/beam)} & \parbox{3.7cm}{$-0.56 \pm 0.03$ \\ ($I < 850$ mJy/beam)} & $-2.03 \pm 0.08$   \\ \hline
        $P \times S$ vs. $a_\mathrm{align}$ & \parbox{3.7cm}{$-5.57 \pm 0.13$ \\ ($I < 350$ mJy/beam)} & \parbox{3.7cm}{$0.12 \pm 0.03$ \\ ($I < 850$ mJy/beam)} & $-0.01 \pm 0.08$ \\ \hline
                
    \end{tabular}
    \label{tab:example_table}
\end{table*}

\setlength{\tabcolsep}{0.6cm}
\renewcommand{\arraystretch}{1}

\begin{table*}[ht]
    \centering
    \caption{Median values of minimum grain alignment and disruption sizes, mean values of magnetic field and median values of magnetic relaxation strengths in the North, Center and South regions.}
    \begin{tabular}{|c|c|c|c|}
        \hline
        
        Parameters & North & Center & South\\ \hline
        $a_\mathrm{align}$ & $0.08 \pm 0.01$ $\mu$m ($T_\mathrm{d} > 19.8$ K) & 0.07 $\mu$m ($T_\mathrm{d} > 25$ K) & $0.09 \pm 0.01$ $\mu$m \\ \hline
        $a_\mathrm{disr}$ & $0.86 \pm 0.11$ $\mu$m ($T_\mathrm{d} > 19.8$ K) & $0.64 \pm 0.05$ $\mu$m ($T_\mathrm{d} > 25$ K) & $1.10 \pm 0.26$ $\mu$m \\ \hline
        $B^*_\mathrm{POS}$ & $100 \pm 40$ $\mu$G & $470 \pm 190$ $\mu$G  & $60 \pm 34$ $\mu$G\\ \hline
        $B^*_\mathrm{tot}$ & $130 \pm 52$ $\mu$G  & $611 \pm 247$ $\mu$G  & $78 \pm 44$ $\mu$G \\ \hline
        Median $\delta_\mathrm{mag,sp}$ & 14.66 & 293.09 & 6.32\\ \hline
        Range of $\delta_\mathrm{mag,sp}$ & $9-25$ & $35-590$ & $5-9$ \\ \hline
     
    \end{tabular}
    
    $^* B_\mathrm{POS}$ values are taken from \cite{2019ApJ...883...95S} and $B_\mathrm{tot}$ values are calculated using relation from \cite{2004ApJ...600..279C}.
    \label{table:magnetic relaxation}
\end{table*}

\section{\bf{Discussions}} \label{section:Discussions}
In this section, we will discuss our results to explain the grain alignment and disruption mechanisms in different sub-regions of the filament with RAT-A and RAT-D.
\subsection{Evidence for the RAT-A mechanism} \label{sec:RAT-A Mechanism}

The decrease in the polarization fraction with the increase in gas column density $N(\mathrm{H_2})$ and total emission intensity $I$, called as polarization hole is found in many observational studies of dust polarization towards different environments (see e.g \citealt{2019FrASS...6...15P, 2021ApJ...908..218H}). The real cause of polarization hole is not yet clear. It can be caused by decrease in grain alignment induced by RATs in denser regions with no embedded sources or by fluctuations in magnetic field orientations possibly induced by turbulence or by both. In G34 filament also which harbors multiple cores in the North and Center regions, the polarization fraction decreases with increasing $I$ and $N(\mathrm{H_2})$ (see Figures \ref{Figure:P_I}, \ref{Figure:P_NH2}). The North and South regions show overall steeper slopes compared to the Center region. However, the Center region has a steep slope of $-0.54 \pm 0.01$ for $I < 850$ mJy/beam which is considered as the regions outside the cores. In our analysis, we find that the polarization fraction decreases with $S$ in the South region and does not show overall significant correlation with $S$ in the North and the Center regions (see Figure \ref{Figure:P_S}). Again, we find that the averaged grain alignment efficiency $P \times S$ decreases with increasing intensity (see Figure \ref{Figure:P_S_I}) in the North and the Center regions more similarly with the variation of polarization fraction with increasing intensity (see Figure \ref{Figure:P_I}) whereas it becomes nearly flat in the South region. The similar trend is also found in the variations of $P$ and $P \times S$ with increasing $N(\mathrm{H_2})$ (see Figures \ref{Figure:P_NH2}, \ref{Figure:P_S_NH2} (left panel), respectively). Therefore, the magnetic field tangling effect is more significant in the South and less significant in the North and the Center regions. The decrease in the averaged grain alignment efficiency with increasing intensity can be explained by the RAT-A Theory.  

Using the RAT-A theory, we calculate the minimum size for aligned grains by RATs using the local physical parameters and obtained a map of alignment size (see Figure \ref{Figure:a_align_map}). We found that the value $a_\mathrm{align}$ is well correlated with the intensity (see Figure \ref{Figure:a_I}) in the North (up to 350 mJy/beam) and South regions which means the alignment size increases in denser regions. But in the North region after an intensity of 350 mJy/beam, the alignment size decreases which could be due to the presence of internal radiation from MM3 core. Also in the Center region, the alignment size increases up to around 850 mJy/beam but decreases at more higher intensity values which could be due to the presence of internal radiation from cores MM1 and MM2. The increasing of alignment size with intensity when no internal radiation source is present is an expectation of RAT-A theory. Also, there is anti-correlation in $P$ with $a_{\mathrm{align}}$ and $P \times S$ with $a_{\mathrm{align}}$ for $I < 350$ mJy/beam (North region) and $I < 850$ mJy/beam (Center region) which are expected to be the regions outside the cores (see Figure \ref{Figure:P_PS_a}). The decrease in grain alignment can be majorly by decrease in RAT alignment efficiency in denser regions in the North except the MM3 core region and the Center regions except the MM1 and MM2 regions. For the South region, there is some effect of magnetic field fluctuations but the decrease in averaged alignment efficiency with $a_{\mathrm{align}}$ can be explained by RAT$-$A theory.

\subsubsection{Correlation of polarization fraction with dust temperature} \label{Section:P_Td}
From the RAT-A theory, it is expected that the polarization fraction $P$ should increase with dust temperature $T_\mathrm{d}$. The alignment size decreases with $T_\mathrm{d}$ and it will give broader size distribution of aligned grains and larger polarization \citep{2020ApJ...896...44L}. We find that the North region shows increase in $P$ with $T_\mathrm{d}$ (see Figure \ref{Figure:P_Td}) up to around 19.8 K and then decreases. Also, in the Center region $P$ increases with $T_\mathrm{d}$ up to around 19.8 K and then decreases. For the South region, $P$ only increases with $T_\mathrm{d}$. Also, $P \times S$ increases with $T_\mathrm{d}$ and then decreases after around 19.8 K as shown by the weighted running means in the North and Center regions but increases overall in the South region (see Figure \ref{Figure:P_S_Td}). We know that the North and Center regions have embedded radiation sources and hence the governing radiation field in these regions is purely diffused ISRF outside MM1, MM2 and MM3 dense cores and is the combination of both the diffused ISRF and the local stellar radiation field in the regions of these dense cores. The South region does not have embedded sources and hence the dust temperature in this region is due to diffused ISRF. The increase in $P$ with $T_\mathrm{d}$ in all the regions can be explained by RAT-A and the decrease in $P$ can be a signature of RAT-D.

\subsection{Implication for RAT-D Mechanism} \label{section:RAT-D Mechanism}
As shown in Figure \ref{Figure:P_Td}, we find the increase of $P$ with increasing $T_\mathrm{d}$ up to nearly 19.8 K and then decreases in the North and the Center regions (more significantly after 25 K in the Center region) which have embedded protostellar cores MM3 (North), MM1 and MM2 (Center). This decreasing of $P$ at larger $T_\mathrm{d}$ values is opposite to the expectation of RAT-A theory (\citealt{2020ApJ...896...44L}). From the variation of averaged grain alignment efficiency $P \times S$ with $T_\mathrm{d}$ (see Figure \ref{Figure:P_S_Td}) it is found that this variation almost nearly follows the variation of $P$ with $T_\mathrm{d}$ (see Figure \ref{Figure:P_Td}). The magnetic field tangling is less significant to cause the depolarization at higher $T_\mathrm{d}$ values in the high $T_\mathrm{d}$ core regions of the North and the Center regions. Also, the median alignment and disruption sizes of grains when $T_\mathrm{d} > 19.8$ K in the North region are nearly 0.08 $\mu$m and 0.86 $\mu$m and when $T_\mathrm{d} > 25$ K in the Center region are nearly 0.07 $\mu$m and 0.64 $\mu$m, respectively. It means that only the grains of sizes ranging from 0.08 $\mu$m to 0.86 $\mu$m can align in the North region of $T_\mathrm{d} > 19.8$ K and only the grains of sizes ranging from 0.07 $\mu$m to 0.64 $\mu$m can align in the Center region of $T_\mathrm{d} > 25$ K. As shown in \cite{1977ApJ...217..425M} the observational upper limit of the grain size distributions in the interstellar medium is $\approx 0.25$ $\mu$m. But in very dense star-forming regions, grain size is expected to increase due to grain growth. \cite{2023ApJ...953...66N} found the implications for grain growth in the G11.11-0.12 filament which has gas volume density in the range from $10^3-10^5$ $\mathrm{cm^{-3}}$, with the lower limit of $a_\mathrm{max}$ greater than 0.3 $\mu$m.    We also expect that the actual size of most of the aligned grains may be far above the estimated $a_\mathrm{align}$ values in very dense regions of the filament with gas volume density in the range of $10^4-10^5$ $\mathrm{cm^{-3}}$ as grain growth is expected in denser regions, thereby further narrowing the grain size distributions of grains that can be aligned. In the very dense and hot core regions of MM1, MM2 and MM3, we can expect high upper limit of $a_\mathrm{max}$ values due to grain growth with the calculated $a_\mathrm{disr}$ values well below these $a_\mathrm{max}$ values. Whereas in the less dense and cold regions outside these cores, the $a_\mathrm{disr}$ values may be higher than the expected smaller upper limit of $a_\mathrm{max}$ values and hence no disruption of grains occurs in these regions. For a fixed upper limit of $a_\mathrm{max}$, when RAT-D occurs larger grains are disrupted thereby removing grains from $a_\mathrm{max}$ to the disruption size $a_\mathrm{disr}$ which results in narrower grain size distribution of grain alignment and hence the polarization fraction gets reduced. The large grains which can be disrupted are the dominant contributors to the observed radiation at 850 $\mu$m.
Therefore, from our calculations of disruption sizes of grains and the observed decrease in polarization fraction in the North and Center regions at larger $T_\mathrm{d}$ values, we can imply the possibility of RAT-D occurring in these regions of the filament due to the presence of internal radiations from the hot and dense cores MM1, MM2 (Center) and MM3 (North). However, the depolarization with higher $T_\mathrm{d}$ values may also be caused by B-field tangling and/or change in the B-field inclination angle within the unresolved beam \citep{2024ApJ...965..183H}. But, there lacks the details on the B-field tangling and inclination variation.

The evidence for the effect of RAT-D was also found in several studies in star-forming regions, e.g Auriga \citep{2021ApJ...908...10N}, Oph A \citep{2021ApJ...906..115T}, BN-KL region \citep{2021ApJ...908..159T, 2024ApJ...974..118N}, M17 \citep{2022ApJ...929...27H}.

\subsection{Role of magnetic relaxation on RAT Alignment} \label{Role of magnetic relaxation}
The importance of enhanced magnetic relaxation combined with RATs to produce perfect grain alignment was first observationally estimated in G11.11-0.12 filament by \cite{2023ApJ...953...66N}. In G34 filament, we find high values of polarization fraction mostly in the outer regions from around 8-20\%. This value of $P$ reaching nearly 20\% is larger than the average value of the diffuse ISM with $P$ nearly $15\%$ observed by \cite{2020A&A...641A..12P}. Only RAT may not be able to produce this high value of $P$ because of its dependence on parameters like grain shape, composition, angle between the direction of radiation field and the magnetic field \citep{2016ApJ...831..159H, 2021ApJ...913...63H}. This high $P$ values in the outer low Stokes $I$ regions may also be caused by missing flux in these regions due to the observation mode. This feature of high P in the outer low Stokes I regions is also observed in several studies using POL-2 and HAWC+ observations (e.g, \citealt{2018ApJ...859....4K, 2018ApJ...861...65S, 2019ApJ...877...88C, 2019ApJ...880...27P, 2019ApJ...876...42W, 2021A&A...647A..78A, 2023ApJ...953...66N}) and in the protostellar envelopes using ALMA observations (e.g, \citealt{2017ApJ...847...92H, 2019ApJ...879...25K}).
However, the missing flux issue was not quantified and studied much. \cite{2020A&A...644A..11L} studied the statistical analysis of dust polarization properties in ALMA observations of class 0 protostellar cores and showed that the interferometric filtering may cause high polarization fraction in the large scale outer low Stokes $I$ regions. However, \cite{2024arXiv240710079C} showed that the high polarization fraction in the outer protostellar envelope observed by ALMA is not entirely due to the interferometric filtering effect and could be produced by the thermal emission of aligned super-paramagnetic grains with $N_\mathrm{cl}$ $\approx$ $10^2-10^3$ which can produce high intrinsic polarization fraction. Apart from possible explanation by missing flux, the observed high $P$ is also possible by the modern grain alignment theory that incorporates the alignment by magnetic relaxation strength. The enhanced magnetic relaxation by grains having embedded iron inclusions can increase the RAT alignment efficiency \citep{2016ApJ...831..159H, 2022AJ....164..248H}. In our estimation of the strength of magnetic relaxation as shown in Figure \ref{Fig:mr_map}, a small level of iron inclusions can produce $\delta_\mathrm{mag,sp} >> 10$, especially in the North and Center regions. The enhanced magnetic relaxation combined with RATs can increase the alignment of grains which may possibly explain the observed high polarization fraction in the outer parts of the North and Center regions of the G34 filament. Therefore, our results further support for the possibility of MRAT alignment mechanism (\citealt{2016ApJ...831..159H}).

\section{\bf{Conclusions}} \label{Section:Conclusions}
In this work, we use thermal dust polarization observations of the G34.43+0.24 filament at 850 $\mu$m using POL-2 instrument mounted on JCMT to study dust grain alignment and disruption mechanisms over all the filament in three sub-regions as North, Center and South. Our main results are summarized as follows:

\vspace{0.3cm}
\textbf{1.} We find decrease in the polarization fraction $P$ with the increase in total intensity $I$ and gas column density $N(\mathrm{H_2})$, so called polarization hole, in all the North, Center and South regions of the filament. We also study the effect of magnetic field tangling on polarization hole by analysing the polarization angle dispersion function. We find that $P$ decreases as $S$ increases in the South region whereas it does not show much significant variations overall in the North and the Center regions. In $P \times S$ variations with $I$ and $N(\mathrm{H_2})$, the slope of South region becomes shallower compared to $P-I$ and $P-N(\mathrm{H_2})$ variations but North and Center regions show steep decrease in averaged grain alignment efficiency almost similar with $P-I$ and $P-N(\mathrm{H_2})$ variations. This suggests that the depolarization in South region has significant contribution from magnetic field fluctuations and for the North and the Center regions, magnetic field tangling does not contribute much in causing depolarization. We find that the decrease in overall averaged grain alignment efficiency with increase in total intensity and gas column density in all the regions can be explained by RAT-A theory except for the MM1, MM2 (Center) and MM3 (North) core regions.

\vspace{0.3cm}
\textbf{2.} To test whether the RAT mechanism can reproduce the observational data, we estimate the minimum grain alignment sizes, $a_{\mathrm{align}}$ for all the regions and the increase in $a_{\mathrm{align}}$ with increasing intensity in the North (up to 350 mJy/beam), Center (upto 850 mJy/beam) and South regions is in agreement with RAT-A theory. Also from the analysis of $P$ and $P \times S$ variation with $a_{\mathrm{align}}$, we find that the effect of magnetic field tangling is significant to cause depolarization in the South region and not much significant in the North and Center regions. The depolarizations in the North and Center regions are mainly due to the decrease in the net alignment efficiency of grains. The decrease in $P$ and $P \times S$ with $a_{\mathrm{align}}$ for the regions except the MM1, MM2 (Center) and MM3 (North) core regions can be explained by the RAT-A Theory.  

\vspace{0.3cm}
\textbf{3.} We find hints for RAT-D effect in the North (MM3) and Center (MM1 and MM2) regions from the analysis of $P$ variation with $T_\mathrm{d}$ and the calculation of minimum alignment and disruption sizes of grains. The variation of $P$ with $T_\mathrm{d}$ is found to be almost similar with the variation of $P \times S$ with $T_\mathrm{d}$ in the North and the Center regions. Our study finds that the decrease in $P$ at large $T_\mathrm{d}$ values in the North and the Center regions can be associated with the regions of the dense and hot protostellar cores MM1, MM2 (Center) and MM3 (North). We find potential evidence for the possible explanation of this observed feature in the core regions by RAT-D mechanism due to the presence of high-luminosity internal radiations from the protostellar cores, although the B-field tangling and B-field inclination angle variations within the unresolved beam can also contribute to the decrease in the polarization fraction.

\vspace{0.3cm}
\textbf{4.} We also studied the importance of enhanced magnetic relaxation on RAT alignment and find the possible scenario that Magnetically enhanced RAT (M-RAT) alignment mechanism can potentially explain the observed high polarization fraction of 8-20\% in the outer protostellar envelopes of the North and the Center regions of the filament, although the high polarization fraction can also be possibly caused by the missing flux in the outer low Stokes $I$ regions. 

\vspace{0.3cm}
In future, we will do pixel by pixel polarization modelling of this G34 region using $DustPol$ and compare with the observational results.

\section{\bf{Acknowledgments}}
This work was partly supported by a grant from the Simons Foundation to IFIRSE, ICISE (916424, N.H.). We would like to thank the ICISE staff for their enthusiastic support. This research has made use of observation data from James Clerk Maxwell Telescope (JCMT) POL-2 instrument. JCMT is operated by the East Asian Observatory on behalf of the National Astronomical Observatory of Japan; Academia Sinica Institute of Astronomy and Astrophysics; the Korea Astronomy and Space Science Institute; the Operation, Maintenance and Upgrading Fund for Astronomical Telescopes and Facility Instruments, budgeted from the Ministry of Finance of China and administrated by the Chinese Academy of Sciences and, the National Key R and D Program of China.

\vspace{0.3cm}
$Software$: Astropy \citep{2013A&A...558A..33A, 2018AJ....156..123A}, Scipy \citep{2020NatMe..17..261V}

\vspace{0.3cm}
$Facilities$: James Clerk Maxwell Telescope (JCMT), Herschel Space Observatory

\appendix

\section{Synthetic intensity map}\label{section:synthetic intensity map}
There may be missing flux in the outer Stokes $I$ low-surface brightness regions of the filament which may produce high polarization fraction.
\begin{figure*}
    \centering
    \includegraphics[width=1\textwidth]{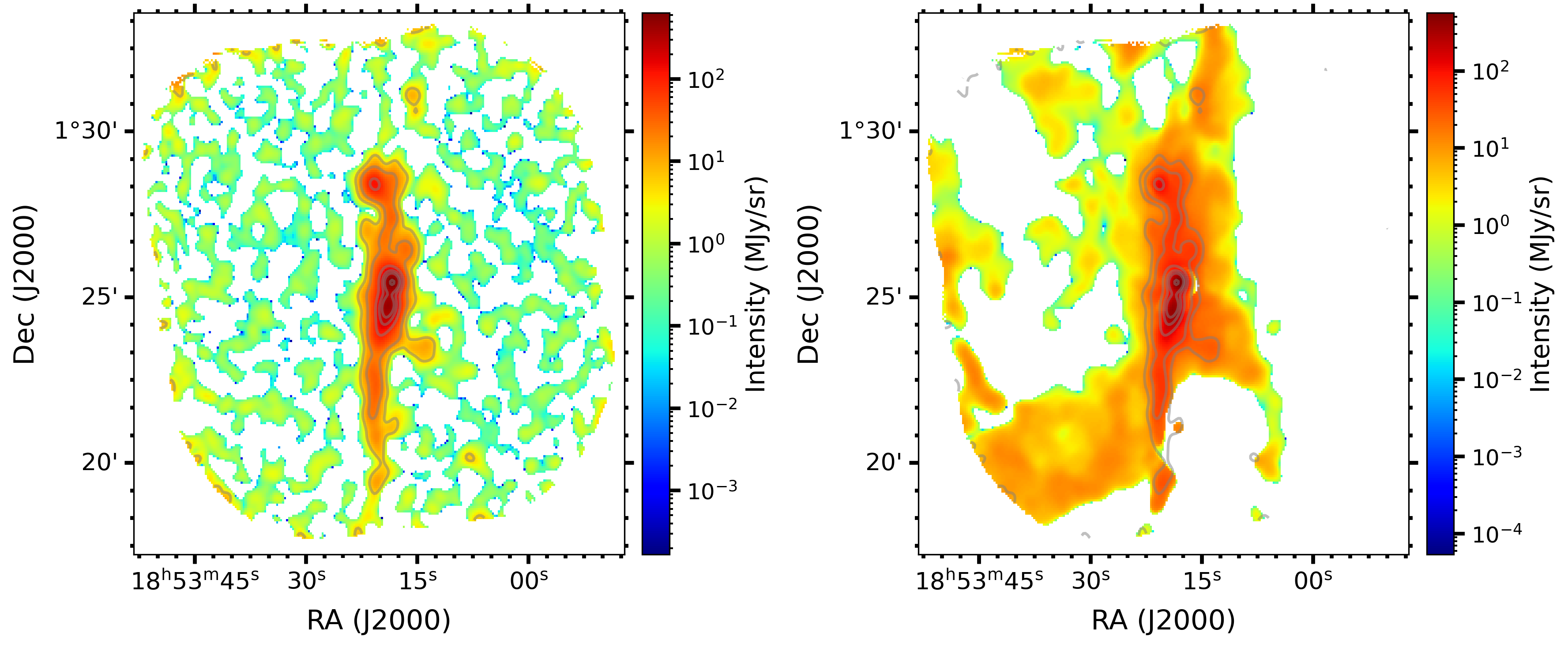}
    \caption{Maps of JCMT/POL-2 observed (left) and synthetic (right) 850 $\mu$m intensities. The contours are drawn at $I$ values of 5, 20, 100, 200, 300, 600 MJy/sr in both the maps. Both the maps are at SPIRE 500 $\mu$m resolution of $35''$.}
    \label{Fig:Observed_Synthetic_map}
\end{figure*}
To check the expected flux in these outer regions, we compute the synthetic 850 $\mu$m intensity map using Herschel PACS/SPIRE data at 70, 160, 250, 350 and 500 $\mu$m. 
\begin{figure*}
    \centering
    \begin{tabular}{ccc}
        \includegraphics[scale=0.25]{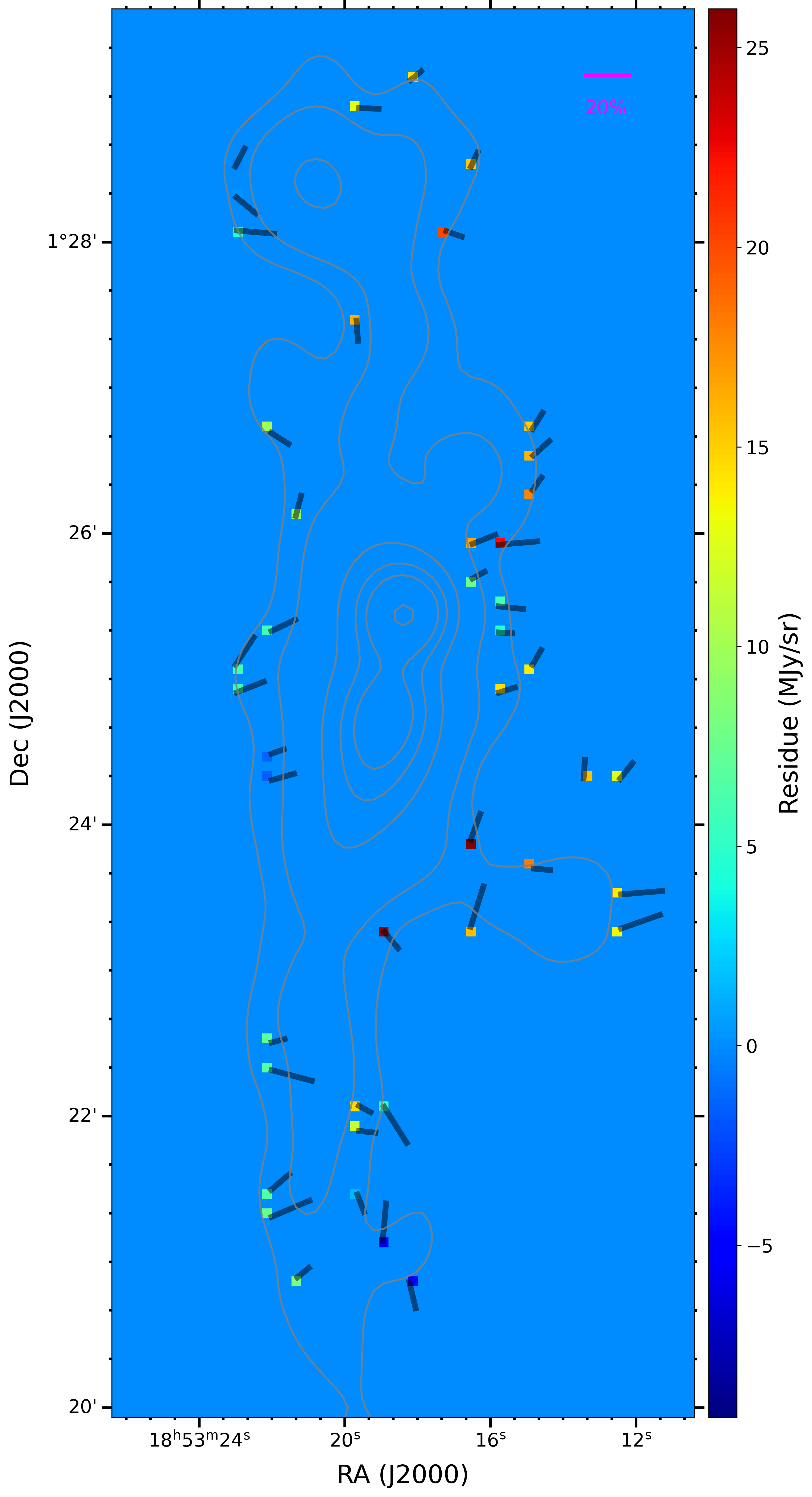} & 
        \hspace{5pt} 
        \includegraphics[scale=0.52]{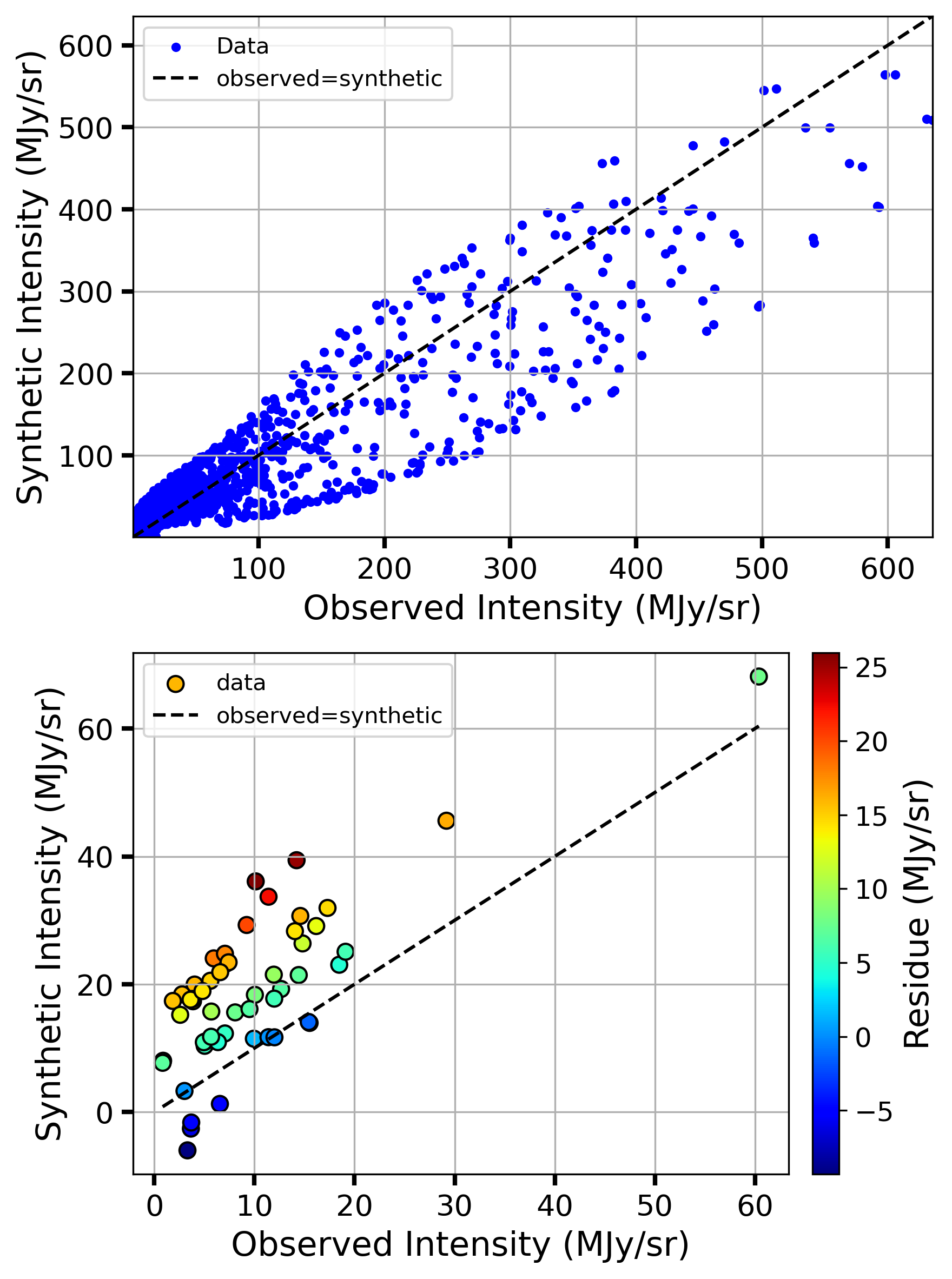} &
    \end{tabular}
    \caption{Map of distribution of residual intensity between the synthetic and observed 850 $\mu$m intensities for the outer regions of the filament (left panel). The gray vectors denote $P$ values ranging from 8-20\%. A reference scale of $P$ is also shown. Right upper panel shows comparison between the observed and synthetic intensities with the dashed black line denoting the equality of both. Right lower panel shows the comparison between the observed and the synthetic intensities only for the outer regions as shown in the left panel with the colorbar denoting the residual intensity values and the dashed black line denoting the equality of both the observed and synthetic intensities.}
    \label{Fig:residue_intensity}
\end{figure*}
The process of estimating the synthetic map is described as follows. First, all the Herschel data at different wavelengths and resolutions are convolved to Herschel 500 $\mu$m FWHM beam size of $35''$ and reprojected on the same grid of the observed JCMT 850 $\mu$m. We use the following modified blackbody function \citep{2008A&A...487..993K} for the spectral energy distribution fitting in each pixel position

\begin{equation}
{
I_{\nu} = B_{\nu}(T_\mathrm{d})(1 - e^{-\tau_{\nu}}),
}
\end{equation}

\begin{equation}
{
B_{\nu}(T_\mathrm{d}) = \frac{2h\nu^3}{c^2} \frac{1}{e^{h\nu/k_\mathrm{B}T_\mathrm{d}} - 1},
}
\end{equation}

\begin{equation}
{
\tau_{\nu} = \mu_{\mathrm{H_2}}m_\mathrm{H}\kappa_{\nu}N(\mathrm{H_2}),
}
\end{equation}
where $B_{\nu}(T_\mathrm{d})$ is the Planck function at the dust temperature $T_\mathrm{d}$, $\tau_{\nu}$ the optical depth, $\mu_\mathrm{H_2}$ the mean molecular weight per hydrogen molecule, $m_\mathrm{H}$ is the hydrogen atom mass, $\kappa_{\nu}$ is the dust opacity and $N(\mathrm{H_2})$ the $\mathrm{H_2}$ column density. We use $\mu_\mathrm{H_2} = 2.8$. We use $\kappa_{\nu}$ values of 1.76, 0.4, 0.195, 0.1, 0.05 $\mathrm{cm^2}$ $\mathrm{g^{-1}}$ \citep{1994A&A...291..943O} considering a dust-to-gas ratio of 0.01 for 70, 160, 250, 350 and 500 $\mu$m, respectively. The best fitted parameters $T_\mathrm{d}$ and $N(\mathrm{H_2})$ are derived for each pixel. Then we use these fitted parameters in the following equation to derive the synthetic 850 $\mu$m intensity map

\begin{equation}
{
I_{850} = B_{850}(T_\mathrm{d})(1 - e^{-\tau_{850}})
}
\end{equation}
We use the dust opacity at 850 $\mu$m, $\kappa_{850}$ value of 0.0197 $\mathrm{cm^2}$ $\mathrm{g^{-1}}$ \citep{1994A&A...291..943O}. The synthetic map thus derived is compared with the observed JCMT/POL-2 850 $\mu$m intensity map convolved to the resolution of SPIRE 500 $\mu$m. We perform a further scaling of the synthetic map to match with the observed intensity by a best-fit scaling factor. The JCMT/POL-2 observed and the synthetic 850 $\mu$m maps (convolved to SPIRE 500 $\mu$m resolution of $35''$) are shown in the left and right panels of Figure \ref{Fig:Observed_Synthetic_map}. Then we compare both the intensities as shown in Figure \ref{Fig:residue_intensity} (right upper panel). For the intensity values below 100 MJy/sr, it seems to be well correlated with some deviations. However, there are much deviations at higher intensity values. We also compute the differences between the synthetic and observed intensities for the outer low-surface brightness regions having polarization fraction in the range from 8-20\% and the map of these differences is shown in Figure \ref{Fig:residue_intensity} (left panel). We find that some regions have somewhat larger difference and some regions small difference. A comparison of the observed and the synthetic intensities for the outer regions is also shown in Figure \ref{Fig:residue_intensity} (right lower panel) with the colorbar denoting the difference values. 
The synthetic intensity map may not be exactly similar to the observed intensity map. Since the intensities have significant deviations in the bright high intensity regions, the expected differences in the outer regions stand only as an approximation which still lacks the accurate comparison to find any missing flux in these outer regions. However, we expect the possibility of having missing flux in the large-scale outer low Stokes $I$ regions.

\bibliography{G34_references}{}
\bibliographystyle{aasjournal}

\end{document}